\documentclass[aps,twocolumn,superscriptaddress,nofootinbib,amsmath,amssymb,floatfix]{revtex4}  
\usepackage{graphicx}  
\usepackage{dcolumn}   
\usepackage{bm,color}        
\usepackage{amssymb}   
\usepackage{adjustbox}
\usepackage{hyperref}

\newcommand{\project}[1]{{\sffamily #1}}

\newcommand{\be}{\begin{equation}}
\newcommand{\ee}{\end{equation}}
\newcommand{\beq}{\begin{equation}}
\newcommand{\eeq}{\end{equation}}
\newcommand{\bea}{\begin{eqnarray}}
\newcommand{\eea}{\end{eqnarray}}

\newcommand{\fnl}{f_{\rm NL}}

\newcommand{\mpl}{m_{\rm Pl}}

\newcommand{\Nps}{N_{\phi_*}}
\newcommand{\Ncs}{N_{\chi_*}}
\newcommand{\Npsps}{N_{\phi_* \phi_*}}
\newcommand{\Ncscs}{N_{\chi_* \chi_*}}

\definecolor{MyBlue}{rgb}{0.15,0.15,0.70}

\hypersetup{
colorlinks=true,
citecolor=red,
linkcolor=MyBlue,
urlcolor=MyBlue
}

\begin{document}
\title{The next non-Gaussianity frontier:\\
{\it what can a measurement with $\sigma(\fnl) \lesssim 1$ tell us about multifield inflation?}}
\author{Roland de Putter} 
\address{California Institute of Technology, MC 367-17, Pasadena, CA 91125, USA}
\author{J\'er\^ome Gleyzes}
\author{Olivier Dor\'e }
\address{California Institute of Technology, MC 367-17, Pasadena, CA 91125, USA}
\address{Jet Propulsion Laboratory, California Institute of Technology, 4800 Oak Grove Drive, Pasadena, CA 91109, USA}

        \date{\today}

\begin{abstract}

Future galaxy surveys promise to probe local primordial non-Gaussianity at
unprecedented precision, $\sigma(\fnl) \lesssim 1$.
We study the implications for multifield inflation by considering spectator models, where
inflation is driven by the inflaton field, but the primordial perturbations are (partially) generated
by a second, spectator field.
We perform an MCMC likelihood analysis using Planck data to study quantitative predictions for $\fnl$ and other observables for a range of such spectator models.
We show that models where the primordial perturbations are dominated by the spectator field, while fine-tuned within the broader parameter space, typically predict $\fnl$ of order unity.
Therefore, upcoming galaxy clustering measurements will constitute a stringent test of whether or not
the generation of primordial perturbations
and the accelerated expansion in the inflationary universe are due to separate phenomena.

\end{abstract}

\maketitle

\section{Introduction}

Inflation is the leading theory describing the early universe \cite{staro80,guth81,linde82,albrechtstein82}, addressing both the standard problems of the hot big bang model (horizon problem, etc) and generating the primordial curvature fluctuations that are the seeds of the structure of the universe observed today.
All this can be achieved, in a manner
 consistent with all current data, with the introduction of a single scalar field.
However, there is no theoretical reason to expect only one field to be important in the early universe, and indeed fundamental physics models, such as those rooted in string theory, commonly predict multiple scalar fields (e.g.~\cite{cvrcekwitten06, dimoetal08, baumc15}).
Moreover, multifield models provide an alternative important for falsifying the single-field paradigm.

A critical
method for testing the single- vs.~multifield nature of inflation is to look for local primordial non-Gaussianity (PNG), characterized by the parameter $\fnl^{\rm loc}$ \cite{komatsuetal05} (since we exclusively consider PNG of the local type, we will often omit the qualifier ``local'').
While single-field models predict negligible local PNG \cite{Maldacena:2002vr,Creminelli:2004yq}, $\fnl = -5/12 \, (n_s - 1)$ (see e.g.~\cite{chenetal13,mooijpalma15} for caveats to this rule), where $n_s$ is the scalar spectral index, multifield models can generate observably large $\fnl$ (see \cite{byrneschoi10} for a review).
The strongest current limits on $\fnl$ come from the bispectra of cosmic microwave background (CMB) fluctuations, observed by the Planck satellite, $\fnl = 0.8 \pm 5.0$ \cite{planck2015png}.
However, near-future galaxy surveys have the potential to improve on this significantly \cite{toronto14}, taking advantage of an exciting new signal.
In the presence of local PNG, the bias of galaxy density perturbations relative to the underlying matter density receives a scale-dependent correction, with scaling $\Delta b(k) \propto k^{-2}$ (where $k$ is the wave number), which becomes important on scales comparable to the Hubble scale \cite{Dalal:2007cu,matver08,slosaretal08,desjseljak10}. By probing this characteristic signal on ultra-large scales, upcoming surveys such as SPHEREx \cite{spherex_wp, Dore:2016tfs,Spherexweb}, LSST \cite{lsst} and EUCLID \cite{euclid} are expected to improve on the current Planck constraint\footnote{Next-generation CMB missions may also significantly improve the $\fnl$ constraint relative to the Planck limit, see e.g.~\cite{S4}.}, eventually
leading to order unity precision \cite{RdPOli14,ferrsmith14, yamauchietal14}, $\sigma(\fnl) \lesssim 1$.

\vskip.1cm

The motivation of this article is to address, in a quantitative way, what such a future $\fnl$ constraint can teach us about multifield inflation, and what signal we might expect to find.

\vskip.1cm

While in single-field models the curvature perturbations are conserved from the time the modes of interest exit the horizon, in multifield inflation the perturbations can undergo super-horizon evolution after this time.
This leads to a rich and complex phenomenology, where the final non-Gaussianity may strongly depend on physics both during and after inflation, including the reheating process.
While large $\fnl$ is not generic in multifield inflation, often requiring significant fine-tuning \cite{Byrnes:2008wi}, a number of interesting models have been identified where large (by {\it large} we mean $|\fnl| \gtrsim 1$) non-Gaussianity is generated.
Examples include the curvaton \cite{lindemukh97,Moroi:2002rd, lythetal03, Lyth:2001nq,Enqvist:2001zp}, modulated reheating \cite{Dvali:2003em,Kofman:2003nx,Bernardeau:2004zz,dvalietal04,zal04}, models with an inhomogeneous end to inflation \cite{bernardeauetal04, lyth05, huang09, kawasakietal09}, the axion-quadratic model where an adiabatic limit is reached before reheating \cite{Elliston:2011dr,leungetal12, ellistonetal14}, two-field models with non-Gaussianity generated during slow-roll inflation \cite{Byrnes:2008wi}, hybrid inflation \cite{byrnesetal09,narukosasaki09}, N-flation \cite{dimoetal08,Kim:2006te,kimetal10}, modulated trapping \cite{langloissorbo09}, and velocity modulation \cite{nakayamasuyama11}.

Although
 many previous studies have provided useful general, analytic insights into generation of non-Gaussianity in multifield inflation (see e.g.~\cite{Byrnes:2008wi,petersontegmark11a,petersontegmark11b,alabidietal10}), an alternative approach is to use MCMC techniques to
numerically sample the full parameter space of a number of multifield models (cf.~\cite{byrnesetal14,smithgrin15, venninetal15,venninetal16}), given constraints from Planck on $\fnl$, $n_s$ and the tensor-to-scalar ratio $r$ (and implicitly on the scalar amplitude $A_s$).
This is the approach we will take here, paying specific attention to the predictions for the distribution of $\fnl$ compared to $\sigma(\fnl) \lesssim 1$.
To compute the evolution of perturbations, and $\fnl$ in particular, we will use the $\delta N$ formalism \cite{staro85,Sasaki:1995aw,lythrodr05,vernwands06}.

Instead of attempting to somehow sample the full space of multifield models, we restrict ourselves to an interesting subset, so-called ``spectator models'' (see also e.g.~\cite{kobaetal13,enqtak13,ellistonetal14,wang16}), that capture
key phenomenology of generating large $\fnl$. As a matter of definition, we assume there are two fields during inflation, described by a separable potential: the {\it inflaton} field, $\phi$, which at horizon exit dominates both the curvature perturbation and the background energy density, and a {\it spectator} field, $\chi$, which is subdominant at this time, but the perturbations of which (partially) determine the final curvature perturbation.
These models are a natural extension beyond single-field inflation, with the inflaton still ``driving'' inflation. In particular, in the ``spectator-dominated'' regime, where the final curvature perturbations are {\it dominated} by the spectator contribution, these models simply separate the two main features of inflation: the inflaton drives the background expansion, while the spectator generates the primordial power spectrum.
Our main interest will be in this latter regime, as it is here that large $\fnl$ can be produced.

In addition to simplifying calculations and thus allowing for easier insights, requiring $\chi$ to be subdominant during inflation (or at least at horizon exit) plays an important role in generating large $\fnl$. The reason is that, quite commonly, if both fields individually have flat potentials (small ratios of the first and second derivatives relative to the potential itself, in Planck units), then $\fnl$ is typically suppressed. However, if the potential of $\chi$ is small compared to the total energy density, it is allowed to have a ``non-flat'' potential while still satisfying the slow-roll conditions (since the latter depend on the ratios of potential derivatives to the {\it total} energy density). As we will see, evading this flat potential restriction is what makes it easier to generate large non-Gaussianity with the spectator field.

Since we wish to sample a concrete parameter space and compare to current data,
we will study three specific spectator models (these models are not necessarily spectator models, but we will restrict them to the parameter space where they are), covering a range of mechanisms for converting perturbations in the spectator field into curvature perturbations: (A) a quadratic-axion potential, with conversion while both fields are rolling, (B) the curvaton, with conversion after the inflaton has decayed into radiation, and (C) modulated reheating, with conversion at the time of reheating.

By choosing specific models, our approach sacrifices generality, but the benefit is that we will be able to derive concrete, quantitative predictions for the probability distribution of $\fnl$ and other quantities. In particular, it is commonly claimed that various models naturally generate order unity $\fnl$ \cite{zal04,suyamaetal13,alabidietal10,renpet15,toronto14}, which we will back up here with a complete likelihood analysis.

Specific questions we will consider are:
\begin{itemize}
\item Do spectator-dominated models (which are the ones interesting for $\fnl$) require fine-tuning of parameters?
\item
What is the posterior probability of finding $|\fnl| \gtrsim 1$ in spectator-dominated models?
\item
How does $\fnl$ relate to model parameters, and what would measuring $\fnl\sim1$ tell us about the parameter space?
\item
What is the complementarity between primordial non-Gaussianity and searches for primordial tensor modes?
\end{itemize}

The article is organized as follows. In Section \ref{sec:form}, we review general formalism describing inflation and the primordial curvature perturbations.
In Section \ref{sec:spec} we introduce spectator models in general and in particular the three scenarios of interest. In Section \ref{sec:obs}, we discuss the observational constraints that we will compare the models with. In Section \ref{sec:results}, we describe the results of our likelihood analysis of the three models and in Section \ref{sec:concl}, we provide a final discussion and summarize our results.

\section{General Formalism}
\label{sec:form}

In single-field inflation,
the comoving curvature perturbation on uniform density hypersurfaces, $\zeta$ \cite{Bardeen:1983qw},
is non-linearly conserved after horizon exit \cite{Wands:2000dp,Weinberg:2003sw}. Thus, the statistics of $\zeta$ remain frozen through
the subsequent evolution of the inflationary universe and the potentially complicated phase of reheating.
In particular, according to the powerful single-field consistency condition \cite{Maldacena:2002vr,Creminelli:2004yq}, the level of local non-Gaussianity remains frozen at the negligible value $\fnl = -5/12 (n_s - 1)$.

The situation in multifield inflation is more complicated. The additional field(s) lead to entropy perturbations at the time of horizon exit, which can be transferred into the curvature perturbation through super-horizon evolution, thus modifying the primordial power spectrum and non-Gaussianity of $\zeta$.
Eventually, inflation ends and reheating takes place, giving rise to the radiation dominated, conventional hot big bang phase. We will assume that here the universe reaches a state of thermal equilibrium without non-local conserved quantum numbers, which implies that after reheating, perturbations are adiabatic, and $\zeta$ is conserved \cite{Weinberg:2004kf,Weinberg:2008si,Meyers:2012ni} until horizon re-entry at a much later time. Thus, the statistics of $\zeta$ just after reheating describe the standard adiabatic primordial fluctuations and feed into the calculation of observational phenomena in the late(r) universe, such as the cosmic microwave background anisotropies and cosmological large-scale structure.

In the models considered in this work, entropy-to-curvature conversion, and in particular the generation of non-Gaussianity in $\zeta$, can take place both during inflation and/or during the period between the end of inflation and the time of reheating. By
the latter we mean the time reheating/thermalization completes and the hot big bang phase with adiabatic perturbations is reached. In the following, we will give a brief overview of our treatment of evolution during both of these phases, and of how the perturbations are computed. We mostly follow standard methods and refer to the vast literature, e.g.~the reviews \cite{byrneschoi10,Wands:2010af}, for more details.

\subsection{Background Evolution}

We will assume inflation is described by two scalar fields with a sum-separable potential,
\beq
W(\phi, \chi) \equiv U(\phi) + V(\chi).
\eeq
The slow-roll parameters are then defined as,
\bea
\label{eq:defslowroll}
\epsilon^\phi &\equiv& \frac{\mpl^2}{2} \, \left(\frac{U_\phi}{W}\right)^2, \quad \eta^\phi \equiv \mpl^2 \, \frac{U_{\phi \phi}}{W} \nonumber \\
\epsilon^\chi &\equiv& \frac{\mpl^2}{2} \, \left(\frac{V_\chi}{W}\right)^2, \quad \eta^\chi \equiv \mpl^2 \, \frac{V_{\chi \chi}}{W},
\eea
where a field subscript defines a derivative w.r.t.~the field and $\mpl = (8 \pi G)^{-1/2}$ is the {\it reduced} Planck mass (the standard Planck mass is $M_{\rm Pl} =\sqrt{8 \pi} \, \mpl$). We work in natural units, $c = \hbar = 1$. We also define a total slow-roll parameter
$\epsilon \equiv \epsilon^\phi + \epsilon^\chi$, which to leading order in slow-roll parameters equals $\epsilon_H \equiv -\dot{H}/H^2$.
Making the standard slow-roll assumption (i.e.~taking the slow-roll parameters above to be small), the Friedmann equation takes the form,
\beq
3 \mpl^2 \, H^2 = U(\phi) + V(\chi),
\eeq
where $H = \dot{a}/a$ is the Hubble rate (dots denote time derivatives).
Under the same slow-roll approximation, the equations of motion for the fields read,
\bea
3 H \dot{\phi} &=& -U_\phi \nonumber \\
3 H \dot{\chi} &=& -V_\chi.
\eea

The above describes the background evolution during inflation while both fields are slowly rolling. After this period, but before reheating, there are many possible scenarios. We will discuss several of these in more detail in the upcoming sections about the three specific models of interest. Broadly speaking, there are two main types of transitions. For a field of mass $m$, when the Hubble rate drops down to $H = m$, the field starts oscillating around the minimum of its potential. Assuming the potential is quadratic around this minimum, the energy density of the field, averaged over oscillation cycles, decays like that of pressureless matter, $\rho \propto a^{-3}$, where $a$ is the cosmic scale factor.

The second type of transition takes place when a field converts its energy into radiation through decays or non-perturbative particle production. This reheating process is generally very complex, but we will model it in a simple way by assuming that reheating occurs instantaneously when $H = \Gamma$, where $\Gamma$ is a decay rate to particles in the post-inflationary heat bath. The instantaneous reheating approximation is commonly made in the literature \cite{Lyth:2001nq,Sasaki:2006kq,Leung:2012ve,Meyers:2013gua}, but it must be noted that a more realistic treatment of reheating could non-negligibly alter the predictions for the primordial perturbations (see \cite{Amin:2014eta} for a review). After this transition, the energy density formerly in the field decays like $\rho \sim a^{-4}$. We will consider scenarios where the two fields are converted to radiation at different times. When used without specific context, we will reserve the term ``reheating'' for the final process leading to the hot big bang phase.

\subsection{Perturbations}

We compute the evolution of perturbations by taking advantage of the separate universe/$\delta N$
formalism \cite{staro85,Sasaki:1995aw,lythrodr05,vernwands06}, which allows us to express the curvature perturbation in terms of field perturbations $\delta \phi_*$, $\delta \chi_*$ at the time of horizon exit, $t_*$. In the large-scale limit, $k \to 0$, we can treat perturbed regions of the universe as separate FLRW universes obeying the background equations. The evolution of perturbations can then simply be obtained by considering the difference between background quantities in these separate universe patches.
In particular, the curvature perturbation $\zeta$ can be computed in terms of the difference in the number of $e$-foldings of evolution between two patches.
Specifically, at $t_*$, on a spatially flat hypersurface, consider a patch of the universe at ${\bf x}$.
Then, the curvature perturbation $\zeta$ at some later time $t_c$ is simply the perturbation to the number of $e$-foldings of expansion needed to
get from $t_*$ to the constant energy density hypersurface, $\delta \rho = 0$, at time $t_c$. In terms of the initial field perturbations on a spatially flat hypersurface, $\delta \phi_* \equiv \phi(t_*, {\bf x}) - \phi(t_*)$, $\delta \chi_* \equiv \chi(t_*, {\bf x}) - \chi(t_*)$ (where $\phi(t_*)$ and $\chi_*$ are the background values),
\bea
\zeta(t_c) &=& \delta N(t_*, t_c) = N_{\phi_*} \, \delta \phi_* + N_{\chi_*} \, \delta \chi_*  \\
&+& \tfrac{1}{2} N_{\phi_* \phi_*} \, \delta \phi_*^2 + \tfrac{1}{2} N_{\chi_* \chi_*} \, \delta \chi_*^2 + N_{\phi_* \chi_*} \, \delta \phi_* \, \delta \chi_* + \dots, \nonumber \label{eqNzeta}
\eea
where we have dropped the position coordinate ${\bf x}$, and on the right hand side also the time dependence.
Here,
\be
N(t_*,t_c) = \int_{*}^c dt \, H(t) \label{defn},
\ee
and $N_{\phi_*} = \partial N/\partial \phi_*$, etc.

The $\delta N$ formalism thus allows us to compute perturbations purely in terms of the evolution of slightly different FLRW background universes. Note that by writing $\zeta$ in terms of $\delta \phi_*$ and $\delta \chi_*$ only, we have implicitly assumed the perturbations
have reached the space of inflationary growing/attractor solutions. In general, a model with two fields has four degrees of freedom, $\delta \dot{\phi}$ and $\delta \dot{\chi}$ in addition to $\delta \phi$ and $\delta \chi$ (defined on a hypersurface of zero spatial curvature). However, by assuming the slow-roll approximation, we have turned second order equations of motion into first order ones, thus reducing the effective number of degrees of freedom to two.

Note finally that at $t_*$, we have to first order,
\bea
\zeta_* = \frac{\sqrt{2 \epsilon^\phi_*}}{2 \epsilon_*} \, \delta \phi_* + \frac{\sqrt{2 \epsilon^\chi_*}}{2 \epsilon_*} \, \delta \chi_*,
\eea
and the non-Gaussianity in $\zeta_*$ is negligible \cite{lidseyseery05}.

\subsection{Connecting to Observation}

In the $\delta N$ formalism, the dimensionless power spectrum of curvature perturbations is given by
\beq
\label{eq:pps}
\mathcal{P}_\zeta = \frac{k^3 \, P_\zeta}{2 \pi^2} = \mathcal{P}_* \, \left( N_{\phi_*}^2 + N_{\chi_*} ^2 \right), \, \text{with} \quad \mathcal{P}_* = \left( \frac{H_*}{2 \pi}\right)^2,
\eeq
where $\mathcal{P}_*$ is the dimensionless power spectrum of the field perturbations $\delta \phi_*$ and $\delta \chi_*$ at horizon exit, and $H_*$ is the Hubble parameter at $t_*$.
A useful quantity in the following is the fraction of the primordial power spectrum generated by the field $\chi$,
\beq
R \equiv \frac{\mathcal{P}_\zeta|_\chi}{\mathcal{P}_\zeta} = \frac{N_{\chi_*}^2}{N_{\phi_*}^2 + N_{\chi_*} ^2}.
\eeq

In addition to the amplitude, another important observational property of the primordial power spectrum
is the spectral index $n_s$. Taking the derivative of the power spectrum (\ref{eq:pps}), one can express this quantity in terms of the slow-roll parameters at $t_*$ \cite{wandsetal02,vernwands06},
\beq
\label{eq:ns def}
n_s - 1 = - 2 \epsilon_* + 2 R \, \eta^\chi_* + 2 (1 - R) \, \eta^\phi_* - \frac{2}{\mpl^2 \, \left( \Nps^2 + \Ncs^2\right)}
\eeq
Similarly, the tensor-to-scalar ratio is given by,
\beq
r = \frac{8}{\mpl^2 \, \left(\Nps^2 + \Ncs^2 \right)}.
\eeq

Finally, the local non-Gaussianity parameter is\footnote{Note that we use a sign convention consistent with Planck, but opposite to that of \cite{Maldacena:2002vr,vernwands06}.},
\beq
\label{eq:fnlgeneral}
\begin{split}
\fnl = \frac56 \bigg[ & (1 - R)^2 \, \frac{\Npsps}{\Nps^2} + R^2 \, \frac{\Ncscs}{\Ncs^2} \\
&+ 2 R \, (1 - R) \, \frac{N_{\phi_* \chi_*}}{\Nps \, \Ncs} \bigg]
\end{split}
\eeq
Thus, non-Gaussianity can be generated through non-linear evolution of initial field perturbations into the curvature perturbation.
The expression above neglects a small, slow-roll suppressed contribution due to intrinsic non-Gaussianity in $\delta \phi_*$ and $\delta \chi_*$ \cite{lidseyseery05}.

\section{Spectator Models}
\label{sec:spec}

As motivated in the Introduction, in this paper we focus on spectator models. We define these here by requiring that the initial curvature perturbation at horizon exit, $\zeta_*$, is dominated by the perturbation in the inflaton $\phi$, so that $\delta \phi_*$ is the initial adiabatic fluctuation. In these models, the perturbation $\delta \chi_*$ therefore describes an entropy perturbation. In equations, this comes down to,
\beq
\label{eq:spec def}
\epsilon^\phi_* \gg \epsilon^\chi_* \quad \& \quad U_* \gg V_* \quad \text{{\bf Spectator Models}} \nonumber
\eeq
In these models, the parameter $R$ now distinguishes between the regimes where the final curvature power spectrum is dominated by inflaton ($\phi$) or spectator ($\chi$) fluctuations,
\bea
R &\approx& 0: \quad \text{{\bf Inflaton-Dominated Regime}} \nonumber \\
R &\approx& 1: \quad \text{{\bf Spectator-Dominated Regime}} \nonumber
\eea
The spectator-dominated regime is particularly interesting for the generation of observable levels of non-Gaussianity and is the main focus of this work.

In spectator models, we generally have
\beq
\label{eq:Nps spec}
\Nps \approx \frac{1}{\sqrt{2 \epsilon_*} \, \mpl} \approx \text{const.},
\eeq
so that the final power spectrum is given by,
\beq
\label{eq:pps spec}
\mathcal{P}_\zeta \approx \frac{1}{1 - R} \, P_{\zeta, *} = \frac{1}{1 - R} \, \frac{1}{2 \epsilon_* \, \mpl^2} \, \left( \frac{H_*}{2 \pi} \right)^2.
\eeq
Thus, the conversion of entropy to curvature can only increase the power spectrum, leading to a large boost in power in the spectator-dominated regime. The scalar spectral index, Eq.~(\ref{eq:ns def}), for spectator models simplifies to,
\beq
\label{eq:ns}
n_s - 1 = -2 \epsilon_* + 2 R \eta^\chi_* + 2 (1 - R) (\eta^\phi_* - 2 \epsilon_*),
\eeq
so that it varies between
\bea
n_s - 1 &=& -6 \epsilon_* + 2 \eta^\phi_* \quad \text{(inflaton-dominated regime)} \nonumber \\
n_s - 1 &=& -2 \epsilon_* + 2 \eta_*^\chi \quad \text{(spectator-dominated regime)} \nonumber
\eea
Note that both asymptotic values are of order slow-roll, and are entirely determined by the slow-roll parameters at horizon exit.

Since the tensor power spectrum remains constant after horizon exit, the tensor-to-scalar ratio for spectator models is suppressed for $R > 0$,
\beq
\label{eq:r spec}
r = 16 (1 - R) \, \epsilon_* .
\eeq
The suppression of $r$ in the spectator-dominated regime is both a feature and a bug. On the one hand, it becomes exceedingly difficult to detect primordial tensor modes observationally. On the other hand, large-field inflaton models that are currently ruled out because they predict a value of $r$ above the observationally allowed range can be put in concordance with the data by adding a spectator field that seeds a large fraction of the primordial power spectrum (we will discuss this further in Section \ref{subsec:obspros}).

The non-Gaussianity in spectator models is approximately given by (cf.~Eq.~(\ref{eq:fnlgeneral})),
\beq
\fnl = \frac{5}{6} R^2 \, \frac{\Ncscs}{\Ncs^2}.
\eeq
The behavior of $\phi$ is very similar to that of single field inflation, so that the $\Npsps/\Nps^2$ term in Eq.~(\ref{eq:fnlgeneral}) is slow-roll suppressed and thus negligible for our purposes (we remind the reader that we are focused on probing $\fnl$ with order unity precision). Moreover, one can check that in the spectator-dominated regime that interests us,  the cross-term proportional to $N_{\phi_* \chi_*}$ is suppressed compared to the $\Ncscs$ term. We will see that the above expression can give rise to large $\fnl$ for $R \sim 1$.

We will discuss this more quantitatively in Section \ref{subsec:obspros}, but we see here already the complementarity between measuring primordial tensor fluctuations and non-Gaussianity (of the scalar fluctuations).
For large-field inflaton potentials in the inflaton-dominated regime, $r$ is within reach of empirical tests, but $\fnl$ is too small to be detected in the near future, whereas in the spectator-dominated regime, $r$ is strongly suppressed and difficult to detect, but $\fnl$ can be within observational reach.

\vskip.4cm

\subsection{Three Specific Spectator Models}

In the following sections, we will introduce the three specific spectator models for which we will study observational predictions and constraints. For the spectator field potential, $V(\chi)$, we will consider both an axion-like periodic potential and a quadratic potential,
which we will discuss in more detail further below.

For the inflaton field, we consider simple, large-field (e.g.~power law) potentials. Since our main interest is in the properties of the spectator field and how it generates $\fnl$, the details of $U(\phi)$ are less important for our study than the properties of $\chi$ and the transfer of its perturbations. The spirit of our approach to the inflaton potential is that there is in principle ample freedom in its form to always be able to fit at least $n_s$ and $A_s$, and that $\fnl$ is relatively insensitive to the details of $U(\phi)$. In practice, to keep the calculations as simple as possible, our main implementation of the inflaton potential will be a quadratic potential,
\beq
U(\phi) = \tfrac{1}{2} m_\phi^2 \, \phi^2,
\eeq
where we treat the field value at horizon exit, $\phi_*$, as a free parameter.
We will briefly consider a more general setup, with varying power law index of the inflaton potential, in Section \ref{subsec:obspros}.

\subsubsection{Case A: The Quadratic-Axion in the Horizon Crossing Approximation}
\label{subsec:intro case A}

We first consider a model where the spectator field is governed by a periodic, axion-like potential,
\beq
\label{eq:axion pot}
V(\chi) = \tfrac{1}{2} V_0 \, \left[ 1 + \cos\left( \frac{2 \pi \chi}{f} \right) \right],
\eeq
where $f$ is a ``decay constant'', and $V_0$ gives the normalization of the potential. In the case under consideration where the inflaton is described by a quadratic potential, this is the quadratic-axion model \cite{Elliston:2011dr,leungetal12, ellistonetal14}. This is a known, simple example of a model capable of generating large non-Gaussianity ($|\fnl| \gtrsim 1$) during or slightly after slow-roll inflation without necessarily relying on mechanisms during the reheating phase. We explain below why this is, after introducing the approximation we will use to compute perturbations in this model. The assumption of a quadratic potential for the inflaton is not crucial so that the quadratic-axion model is merely a specific example of a broader class of ``inflaton-axion'' models.

During inflation, while both fields obey the slow-roll conditions, the number of $e$-foldings between $t_*$ and some later time $t_c$, is given by\footnote{The case of a sum-separable potential during slow-roll is special in the sense that $N$ can be written as a path-independent integral through field space. In other words, there exists some function defined for all $\phi$ and $\chi$, and $N$ is simply the difference of that function between the end point $(\phi_c, \chi_c)$ and the starting point $(\phi_*, \chi_*)$.
It is this property that allows the derivation of closed analytic expressions for the curvature perturbation as in \cite{vernwands06}.}
\beq
N = -\frac{1}{\mpl^2} \, \int_{\phi_*}^{\phi_c} \, \frac{U}{U_\phi} \, d\phi - \frac{1}{\mpl^2} \, \int_{\chi_*}^{\chi_c} \, \frac{V}{V_\chi} \, d\chi.
\eeq
Assuming a zero spatial curvature hypersurface at $t_*$ and a constant energy density surface at $t_c$, the $\delta N$ formalism allows us to write the curvature perturbation at $t_c$ as,
\bea
\zeta(t_c) &=& \delta N = \left[ \frac{1}{\mpl^2} \, \left( \frac{U}{U_\phi} \right)_* \, \delta \phi_* + \frac{1}{\mpl^2} \, \left( \frac{V}{V_\chi} \right)_* \, \delta \chi_* \right] \nonumber \\
&-&  \left[ \frac{1}{\mpl^2} \, \left( \frac{U}{U_\phi} \right)_c \, \delta \phi_c + \frac{1}{\mpl^2} \, \left( \frac{V}{V_\chi} \right)_c \, \delta \chi_c \right],
\eea
which is straightforwardly extended to higher orders.
Since we have fixed the gauge at $t_c$, the perturbations $\delta \phi_c$ and $\delta \chi_c$ are fully specified in terms of $\delta \phi_*$ and $\delta \chi_*$. This enabled \cite{vernwands06} to derive analytic expressions for $\zeta$ in terms of $\delta \phi_*$ and $\delta \chi_*$ only. The $\delta \phi_c$ and $\delta \chi_c$ contributions make these expressions rather complicated.

{\em The Horizon Crossing Approximation (HCA) -}
If, however, before $t_c$, an adiabatic limit is reached where, independently of the initial perturbation, the fields always end up on the same field trajectory, the contributions from the perturbations at $t_c$ can be neglected. In this limit, the perturbations are well described by the so-called Horizon Crossing Approximation (HCA) \cite{alabidilyth06, Kim:2006te, Elliston:2011dr} and are fully expressed in terms of the field perturbations at horizon exit (we note that, in single-field inflation, this assumption is generally satisfied at all times after horizon exit under the standard assumption of being on the single-field attractor solution, thus explaining why $\zeta$ is conserved in single-field inflation).
The HCA simplifies the expressions for the perturbations and their non-Gaussianity considerably, giving easy insights in the multifield phenomenology and allowing us to straightforwardly identify models with the potential for generating large non-Gaussianity.

Before explicitly writing the HCA expressions to second order, it is useful to define slow-roll parameters for the individual potentials (cf.~Eq.~(\ref{eq:defslowroll})),
\bea
\tilde{\epsilon}^\phi &\equiv& \frac{\mpl^2}{2} \, \left(\frac{U_\phi}{U}\right)^2, \quad \tilde{\eta}^\phi \equiv \mpl^2 \, \frac{U_{\phi \phi}}{U} \\
\tilde{\epsilon}^\chi &\equiv& \frac{\mpl^2}{2} \, \left(\frac{V_\chi}{V}\right)^2, \quad \tilde{\eta}^\chi \equiv \mpl^2 \, \frac{V_{\chi \chi}}{V}. \nonumber
\eea
While the true slow-roll parameters, normalized by the total energy density $W$, are required to be small for the slow-roll approximations to hold, the individual slow-roll parameters can in principle be larger than unity. In particular, for spectator models,
$\tilde{\epsilon}^\phi_* \approx \epsilon^\phi_* \approx \epsilon_*$, $\tilde{\eta}^\phi_* \approx \eta^\phi_*$, but $\tilde{\epsilon}^\chi_* = (W_*/V_*)^2 \, \epsilon^\chi_* \gg \epsilon^\chi_*$ , $\tilde{\eta^\chi}_* = (W_*/V_*) \, \eta^\chi_* \gg \eta^\chi_*$.

In terms of these, the HCA gives,
\bea
\label{eq:NHCA}
\mpl \, \Nps = \frac{1}{\sqrt{2 \tilde{\epsilon}^\phi_*}} &,& \quad \mpl \, \Ncs = \frac{1}{\sqrt{2 \tilde{\epsilon}^\chi_*}}  \\
\mpl^2 \, \Npsps = 1 - \frac{\tilde{\eta}^\phi_*}{2 \tilde{\epsilon}^\phi_*} &,& \quad
\mpl^2 \, \Ncscs = 1 - \frac{\tilde{\eta}^\chi_*}{2 \tilde{\epsilon}^\chi_*},\nonumber
\eea
and $N_{\phi_* \chi_*}=0$. Thus, assuming a spectator model, to be in the spectator-dominated regime, say $\Ncs^2 > \Nps^2$, one requires
\beq
\tilde{\epsilon}^\chi_* < \tilde{\epsilon}^\phi_*\simeq \epsilon_*.
\eeq
This means one needs a very small value of $\epsilon^\chi_*$. Next, assuming the spectator domination requirement is fully satisfied ($R \approx 1$), the non-Gaussianity is given by,
\beq
\label{eq:fnlHCA}
\fnl \sim \frac{5}{6} \, \frac{\Ncscs}{\Ncs^2} = \frac{5}{6} \, \left(2 \tilde{\epsilon}^\chi_* - \tilde{\eta}^\chi_* \right) \approx - \frac{5}{6} \, \tilde{\eta}^\chi_*.
\eeq
Therefore, for a spectator dominated model to generate large non-Gaussianity, one needs a large individual slow-roll parameter $|\tilde{\eta}^\chi_*| \gtrsim 1$. This is not inconsistent with slow-roll inflation because $\eta^\chi_*$ is suppressed relative to $\tilde{\eta}^\chi_*$. This argument (based on the simple HCA assumption), nicely illustrates the more general point that, in multifield inflation, if both fields contribute significantly to the energy density of the universe, the slow-roll conditions typically restrict $\fnl$ to be small, and that this limitation can be evaded by considering spectator fields, which may have very non-flat potentials without violating slow-roll because their energy density contribution is small.

\begin{figure}[h!]
\centering
\includegraphics[width=.35\textwidth]{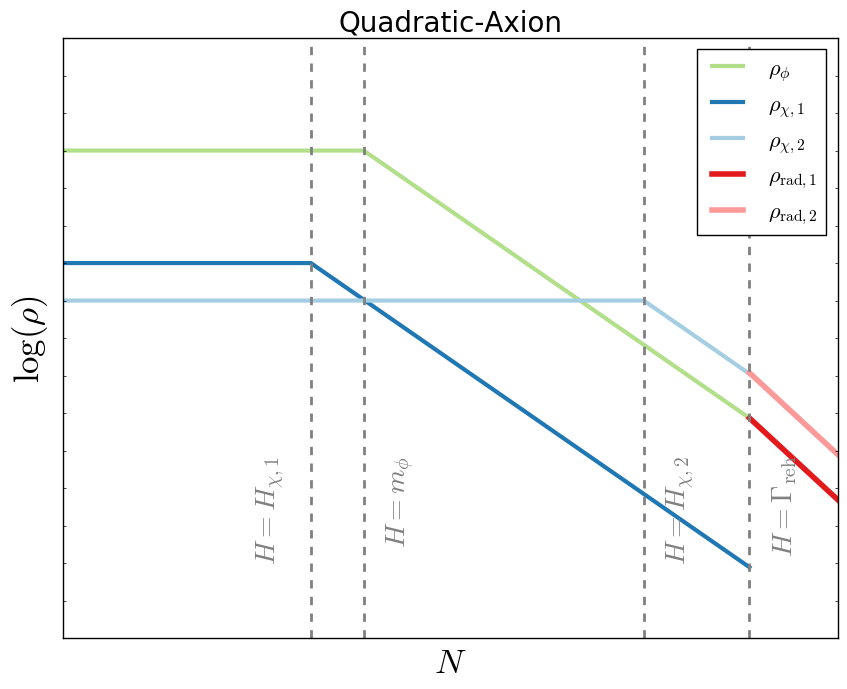}
\caption{Illustration of the behavior of the background densities in the quadratic-axion scenario (Case A), as a function of the number of $e$-folds $N$. We show two scenarios, depending on whether the spectator field/axion, $\chi$, starts rolling down its potential before (dark blue) or after (light blue) the end of inflation.
As illustrated, in the latter case, $\chi$ could generate a brief second phase of inflation
before it starts oscillating.
We assume that reheating only occurs after the phase is reached where both energy densities decay like $\rho \sim a^{-3}$ and that the reheating process does not alter the curvature perturbations. After reheating,
the universe is filled with radiation (thick red and pink for scenario 1 and 2 respectively) with adiabatic perturbations.}
\label{fig:QAdens}
\end{figure}

The requirements of ultra-small $\tilde{\epsilon}^\chi$ and large $\tilde{\eta}^\chi$ are naturally incorporated in the axion model, Eq.~(\ref{eq:axion pot}), if the initial field $\chi_*$ is placed near the top of the cosine potential. In the limit $\chi_*/f \to 0$, the slope of the potential asymptotes to zero, while the curvature approaches a constant, thus satisfying the two conditions. By having the amplitude $V_0$ low, the slow-roll conditions are satisfied as well. This ability to produce large $\fnl$, together with the fact that axion potentials can be realized in a technically natural way from a more complete Lagrangian, makes the inflaton-axion model theoretically appealing.

To illustrate the background evolution of the fields in this model, we schematically plot the energy densities as a function of the number of $e$-folds $N$ in Figure \ref{fig:QAdens}.
We show two scenarios.
In both, the energy density of $\chi$ is subdominant throughout the inflationary period driven by the inflaton $\phi$. In the first scenario, $\chi$ starts rolling with its energy density decaying according to $\rho_\chi \propto a^{-3}$ slightly before the end of inflation (dark blue curve). After this, inflation ends, and both components decay like matter. Later, reheating takes place, after which we assume the total energy density of the universe to exist in the form of radiation (thick red curve).
In the alternative scenario, $\chi$ starts rolling/oscillating after the end of inflation. While not always the case, in the scenario shown, this happens {\it after} $\chi$ has come to dominate the energy budget of the universe, thus leading to a second phase of inflation of modest duration. Again, after both fields end up decaying proportional to $a^{-3}$, reheating takes place, and the universe is filled with radiation (thick pink).

Which scenario takes place depends on the model parameters in a relatively straightforward manner.
The time that $\chi$ starts rolling (exits slow-roll) is partially determined by comparing the Hubble rate to the mass associated with the axion potential,
\beq
\label{mchiV0}
m^2_\chi \equiv \frac{2 \pi^2}{f^2} \, V_0.
\eeq
Tuning the initial field value to be close to the hilltop, $\chi_*/f \ll 1$, however, will delay this moment.
For $m_\chi > m_\phi$ and $\chi_*/f$ not too small, $\chi$ can thus start rolling before the end of inflation, as shown in scenario 1.
If $V_0$ is sufficiently large, $\chi$ can also come to dominate the universe before the end of inflation, thus lengthening the duration of inflation (not shown).
In most cases relevant to our likelihood analysis, $\chi$ starts oscillating well after the end of inflation.
For large $f$ and small $\chi_*/f$,
$\chi$ first drives a second phase of inflation (scenario 2), but in a large fraction of parameter space, this is not the case, i.e.~$\chi$ starts oscillating when its energy density is smaller than or comparable to $\rho_\phi$ (not shown).

{\em Validity of Horizon Crossing Approximation -}
We now come back to the question of the range of validity of the HCA. The approximation is exact if an adiabatic limit is reached while both fields are in the slow-roll regime\footnote{When this is not the case, it is possible to modify $\zeta$ at the end of inflation through the dependence on $\delta \phi_c$
and $\delta \chi_c$ which is neglected in the HCA. In particular, models where the fields are on a turning trajectory at the end of inflation can generate large $\fnl$ in a way not captured by the HCA.}.
In practice, even if this is not the case, if after inflation a phase is reached where both fields oscillate around their minima (with energy densities decaying like $\rho \propto a^{-3}$), so that $\zeta = \text{const.}$\footnote{While $\zeta$ is constant in such a phase, this is not necessarily an adiabatic limit, as entropy perturbations may still exist. Only if these entropy perturbations are not converted to curvature through reheating at a later stage, will $\zeta$ remain constant into the hot big bang phase.}, then the HCA still turns out to be a reasonable approximation in many cases (see e.g.~\cite{Elliston:2011dr}).

To test the range of validity of the HCA in the quadratic-axion model, we have numerically computed the perturbations into the $\zeta = \text{const.}$ phase using the exact $\delta N$ formalism and compared the results to the HCA predictions. We describe the details in Appendix \ref{app:HCA}, but the main result is that, for the parameter space we will study here, the Horizon Crossing Approximation is a good estimator of $\fnl$ to within a factor of less than two. In addition, we introduce an $f$-dependent correction factor that brings the HCA prediction in much better agreement with the exact numerical calculation. We use both prescriptions separately in our likelihood analysis to bracket the possible range of $\fnl$ values. Both prescriptions give qualitatively similar results.

Finally, for Case A, we assume that reheating does not modify $\zeta$ after the $\zeta = \text{const.}$ phase described by the HCA. In the simplified instantaneous reheating picture, this would correspond to reheating taking place on a constant total energy density hypersurface.

\subsubsection{Case B: The Curvaton}

For Case B, we consider a simple quadratic potential for the spectator field,
\beq
V(\chi) = \tfrac{1}{2} m_\chi^2 \, \chi^2.
\eeq
The curvaton scenario \cite{lindemukh97,Moroi:2002rd, lythetal03, Lyth:2001nq,Enqvist:2001zp} relies on a post-inflationary phase where $\phi$ has already decayed into radiation and $\chi$ is oscillating around its minimum. Thus, the energy density of $\chi$ grows relative to that of $\phi$ and perturbations $\delta \chi$ are converted into curvature perturbations. It is known that in the limit where the curvature perturbations are dominated by $\delta \chi$, large non-Gaussianity ($|\fnl| \gtrsim 1$) can be generated \cite{Sasaki:2006kq}.

\begin{figure}[h!]
\centering
\includegraphics[width=.35\textwidth]{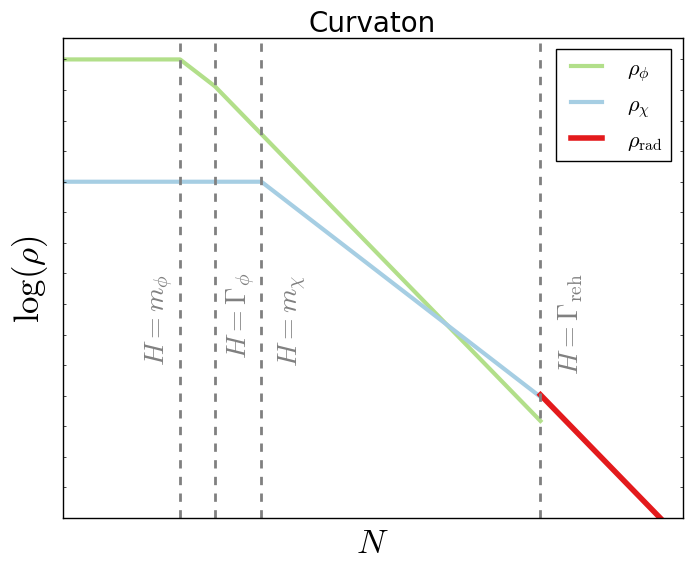}
\caption{Illustration of the behavior of the background densities in the curvaton scenario (Case B).
After the inflaton has decayed into radiation (at $H=\Gamma_\phi$), $\rho_\phi \sim a^{-4}$, an until then subdominant spectator/curvaton field $\chi$ starts oscillating around its potential minimum. Since $\rho_\chi \sim a^{-3}$ in this phase, its energy density may become important and its perturbations can be converted into curvature perturbations. This curvaton phase ends at $H=\Gamma_{\rm reh}$, when we assume $\chi$ decays into radiation with adiabatic fluctuations without further modifying the curvature perturbations.}
\label{fig:curvdens}
\end{figure}

Here, we consider the following specific curvaton scenario, with three main phases, as illustrated in Fig.~\ref{fig:curvdens}. The first phase is the period of inflation, where both fields are slowly rolling. This phase ends when $H_{\rm end} = m_\phi$, at $t_{\rm end}$, after which the inflaton starts oscillating around its minimum, with energy density decaying like pressureless dust, $\rho_\phi \propto a^{-3}$. We assume that at some point during this phase, the inflaton decays into radiation, leading to $\rho_\phi \propto a^{-4}$ evolution (we keep using the subscript $\phi$ even though at this point the component consists of radiation). The second phase ends at $H_{\rm curv} = m_\chi < m_\phi$, at $t_{\rm curv}$, when the spectator field starts oscillating around its minimum, leading to $\rho_\chi = \tfrac{1}{2} m_\chi^2 \, \chi_{\rm curv}^2 \, (a/a_{\rm curv})^{-3}$. We refer to this third phase as the curvaton phase. It ends when also the curvaton decays into radiation at $H_{\rm reh} = \Gamma_{\rm reh} < m_\chi$. We assume all transitions take place on constant total energy density slices so that $\zeta$ is conserved across the transitions. While we assume throughout this paper that after reheating the perturbations are purely adiabatic, we refer to \cite{smithgrin15} for a recent study of the observational consequences of persisting isocurvature fluctuations.

For the curvaton scenario, we will make a slightly stronger assumption than the usual spectator requirements, namely that $\chi$ is subdominant not just at $t_*$, but until the beginning of the curvaton phase, $t_{\rm curv}$, i.e.~$\tfrac{1}{2} m_\chi^2 \, \chi_{\rm curv}^2 \ll 3 \mpl^2 \, m_\chi^2$. This allows for simple analytic expressions for the spectator contributions to the final curvature perturbations \cite{Sasaki:2006kq} and for the evolution from $\chi_*$ to $\chi_{\rm curv}$. At the linear level, we use
\beq
\label{eq:Ncs curv}
\Ncs = \frac{2 r_{\chi,{\rm reh}}}{3 \chi_*},
\eeq
where $\chi_*$ is the initial field value and
\beq
r_{\chi} \equiv \frac{3 \rho_{\chi}}{3 \rho_{\chi} + 4 \rho_{\phi}}
\eeq
gives the relative contribution of the curvaton to $\rho + 3 p$ during the curvaton phase ($r_{\chi,{\rm reh}}   = r_\chi$ evaluated at $t_{\rm reh}$).
The non-Gaussianity parameter is given by
\beq
\fnl = \frac{5}{6} R^2 \, \left( -r_{\chi,{\rm reh}} - 2 + \frac{3}{2 r_{\chi,{\rm reh}}} \right).
\eeq
From Eq.~(\ref{eq:Ncs curv}), one needs small $\chi_*$ ($\lesssim \sqrt{2 \epsilon_*} \, \mpl$) to reach the spectator-dominated regime. Assuming $R \sim 1$ is indeed obtained during the curvaton phase, if this happens while $r_\chi$ is small, the non-Gaussianity can be very large $\fnl \sim 5/(4 r_{\chi})$. If and when the curvaton phase continues to the point where $\rho_\chi$ dominates ($r_\chi \to 1$), the asymptotic value $\fnl = -5/4$ is reached.

The results only minimally depend on exactly when during the second phase the inflaton decays into radiation\footnote{Varying the time of decay slightly changes the evolution of $\chi$ between $t_{\rm end}$ and $t_{\rm curv}$ and therefore affects the curvaton energy density at $t_{\rm curv}$ and thus $r_\chi$.}. Therefore, to keep the analysis minimal, instead of including a free parameter $\Gamma_\phi$ to describe this transition, we simply consider the extreme case, where $\phi$ decays immediately at $t_{\rm end}$. We have checked that using the opposite extreme, where it decays at $t_{\rm curv}$, leads to very similar results.

\subsubsection{Case C: Modulated Reheating}

For Case C, we again consider a simple quadratic potential for the spectator field,
\beq
V(\chi) = \tfrac{1}{2} m_\chi^2 \, \chi^2.
\eeq
In the modulated reheating scenario \cite{Dvali:2003em,Kofman:2003nx,Bernardeau:2004zz,dvalietal04,zal04} (see, e.g.~\cite{kobayashietal14,meyerstarrant14} for recent studies), the decay rate of the inflaton, which determines the time of reheating, depends on the spectator field $\chi$. Then, even if $\chi$ contributes negligibly to the energy density of the universe, the quantum fluctuations in $\chi$ at horizon exit can be transferred into curvature perturbations through the reheating process (the reheating hypersurface is not one of constant energy density, but is modulated by $\chi$).
This is a well known scenario producing large $\fnl$ \cite{zal04}.

\begin{figure}[h!]
\centering
\includegraphics[width=.35\textwidth]{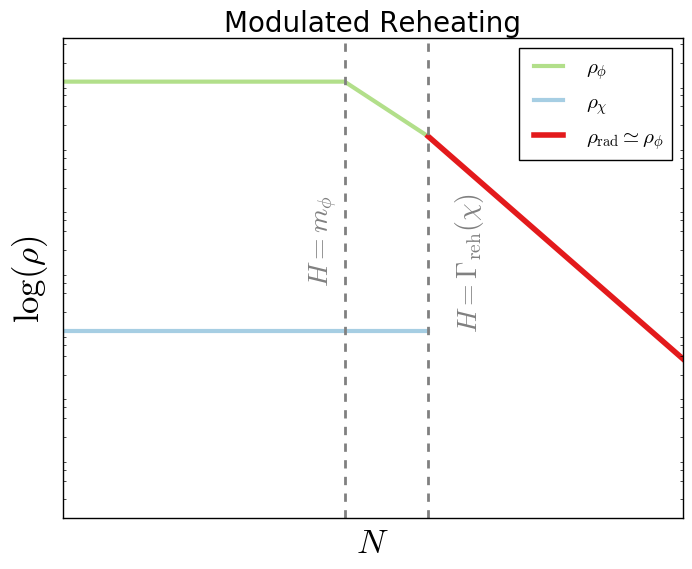}
\caption{Illustration of the behavior of the background densities in the modulated reheating scenario  (Case C).
After inflation, the inflaton oscillates around its potential minimum and decays into radiation when $H = \Gamma_{\rm reh}(\chi)$. The spectator field $\chi$, which has negligible energy density at the time of reheating, modulates the hypersurface of reheating and may thus convert its fluctuations into curvature perturbations. We assume that after reheating by $\phi$, the spectator field $\chi$ plays no further role (e.g.~promptly decaying itself), leading to a universe composed of radiation with adiabatic perturbations.}
\label{fig:MRdens}
\end{figure}

The specific case we consider here, see Figure \ref{fig:MRdens}, consists of two phases: the standard inflationary phase, ending at $H_{\rm end} = m_\phi$, and a subsequent phase where the inflaton oscillates around its minimum and $\chi$ is still slowly rolling. This phase ends at $t_{\rm reh}$, when $H_{\rm reh} = \Gamma_{\rm reh}(\chi)$. We assume $\chi$ is subdominant all the way up to $t_{\rm reh}$, $\tfrac{1}{2} m_\chi^2 \, \chi_{\rm reh}^2 \ll 3 \Gamma_{\rm reh}^2(\chi_{\rm reh})$, and that it plays no role in the generation of curvature perturbations other than through the reheating process. In particular, we assume that after $t_{\rm reh}$, $\chi$ decays into radiation as well without further modifying $\zeta$.

For the reheating of the inflaton, we consider a toy model where decay to fermions ($q$) is the dominant process, through a coupling term of the form,
\beq
\label{eq:coupling}
\mathcal{L} \supset - \lambda(\chi) \, \phi \, \bar{q} \, q.
\eeq
The decay rate is then \cite{lindebook05}
\beq
\Gamma_{\rm reh} = \frac{m_\phi \, \lambda^2(\chi)}{8 \pi}.
\eeq
For the dependence of the coupling constant on $\chi$, we choose a simple expansion truncated at quadratic order (see also, e.g., \cite{ichikawaetal08, kamadaetal11}),
\beq
\label{eq:lambda def}
\lambda(\chi) = \lambda_0 + \lambda_1 \, \frac{\chi}{M_c} + \tfrac{1}{2} \lambda_2 \, \left( \frac{\chi}{M_c} \right)^2,
\eeq
where $M_c$ is a cutoff scale in the effective field theory, $M_c \gg H_*$, and the dimensionless parameters $\lambda_0, \lambda_1, \lambda_2$ are at most of order unity (additional bounds are described in Appendix \ref{app:priors}).

Following \cite{kobayashietal14}, we use,
\beq
\label{eq:Ncs MR}
\Ncs = -\frac{1}{6} \, \frac{\Gamma'_{\rm reh}}{\Gamma_{\rm reh}} \, \frac{\partial \chi_{\rm reh}}{\partial \chi_*},
\eeq
and
\beq
\label{eq:fnl MR}
\fnl = 5 \left[  1 - \frac{\Gamma_{\rm reh} \, \Gamma''_{\rm reh}}{\left( \Gamma'_{\rm reh}\right)^2}
- \frac{\Gamma_{\rm reh}}{\Gamma'_{\rm reh}} \, \left( \frac{\partial \chi_{\rm reh}}{\partial \chi_*} \right)^{-2}
\, \frac{\partial^2 \chi_{\rm reh}}{\partial \chi_*^2} \right],
\eeq
where primes denote derivatives w.r.t.~$\chi_{\rm reh}$.
We include the contributions due to the evolution of the spectator field between $t_*$ and $t_{\rm reh}$ as in \cite{kobayashietal14}.
The expression for $\fnl$ shows that, if the spectator-dominated regime is reached, one would naturally expect $|\fnl| \sim 5$. We will make this more quantitative in Section \ref{subsec:MR results}.

\section{Comparison to Observation}
\label{sec:obs}

\subsection{Current CMB Constraints}

We derive constraints on the three spectator models discussed above using the most recent Planck cosmic microwave background measurements \cite{planck2015} of $n_s$ and $\fnl$. For $r$, we use the joint analysis by Planck and BICEP2 of B-modes on the subset of the sky covered by BICEP2 \cite{Ade:2015tva} (which is why we will not include a correlation between $n_s$ and $r$ in our likelihood). Specifically, we model the measurements by Gaussian likelihoods (restricted to $r \ge 0$ for $r$) with mean and standard deviation,
\beq
n_s = 0.9645 \pm 0.0049,
\eeq
\beq
r = 0.0497 \pm 0.0383,
\eeq
where $n_s$ and $r$ are defined relative to a pivot scale $k_* = 0.05 \, $Mpc$^{-1}$, and
\beq
\fnl = 0.8 \pm 5.0.
\eeq
We then apply standard Markov Chain Monte Carlo (MCMC) techniques using the python package \project{emcee} \cite{ForemanMackey:2012ig} to derive constraints on the spectator parameters.

We summarize the parameter space and physically motivated priors for each model in Appendix \ref{app:priors}.

Planck also provides a measurement of the amplitude of the primordial power spectrum, $A_s \equiv \mathcal{P}_\zeta(k_*)$, namely \cite{planck2015, Ade:2015xua}
\beq
\ln\left( 10^{10} A_s \right) = 3.094 \pm 0.034.
\eeq
We treat this measurement differently than the constraints on $n_s$, $r$ and $\fnl$.
The reason is that we still have the freedom to take out an absolute energy scale from the equations describing our models by rescaling various model parameters. We can choose this energy scale to be the normalization of the inflaton potential, in this case $m_\phi^2$. Specifically, if we define rescaled quantities, $V(\chi) \to \tilde{V}(\chi) \equiv V(\chi)/m_\phi^2$, $m \to \tilde{m} \equiv m/m_\phi$, $\Gamma \to \tilde{\Gamma} \equiv \Gamma/m_\phi$, etc. (but leave the fields unchanged), the evolution equations retain the same form given previously, but in terms of the ``tilded'' quantities.
The observables $n_s, r$ and $\fnl$ are also independent of the overall mass scale $m_\phi$.
The main quantity that does depends on $m_\phi$ is $A_s$. Therefore, we will in practice sample the rescaled parameters and, instead of also treating $m_\phi$ as a free parameter, it is implicit that at each point in parameter space it is tuned in order to obey the $A_s$ constraint.
One subtlety in this approach is that physical constraints and priors (see Appendix \ref{app:priors}) sometimes are naturally given in terms of absolute, not rescaled, scales. In order to translate these priors to the rescaled parameters, we will simply use a fiducial value for the overall mass scale, $m_\phi^{\rm fid} = 1.6 \cdot 10^{13} $GeV (the mass scale required to reproduce the observed $A_s$ for an inflaton-dominated model with $\phi_* = 15 \, \mpl$). This is a reasonable choice because the variation in $m_\phi$ needed to fit $A_s$ is relatively small compared to the very wide prior ranges considered here.

In principle, there is a constraint in addition to the measurements of $n_s$, $r$, $\fnl$ and $A_s$, namely on the number of $e$-foldings of inflation between horizon exit of the mode of interest and the end of inflation, $N_*$.
Working backwards in time from the present, one can compute how far outside the horizon a given mode with wave vector $k_*$
was at the time when inflation ends, which in turn specifies how many $e$-folds before the end of inflation that mode must have exited the horizon (see e.g.~\cite{Lyth:1998xn,planck2015}),
\bea
&N_*& + \ln\left( \frac{H_{\rm end}}{H_*}\right) =
61.7 - \ln\left(\frac{k_*}{0.05 \, \text{Mpc}^{-1}}\right) \nonumber \\
&-& \frac{1}{12} \ln\left( \frac{g_*(T_{\rm reh})}{106.75} \right)
+ \frac{1}{4} \ln\left( \frac{H_{\rm end}^2}{3 \mpl^2} \right) \nonumber \\
&+& \frac{1 - 3 w_{\rm reh}}{12 (1 + w_{\rm reh})} \, \ln\left( \frac{H_{\rm reh}^2}{H_{\rm end}^2} \right).
\eea
Here, $w_{\rm reh}$ is the effective equation of state between the end of inflation and the finalization of the reheating phase, and $g_*(T_{\rm reh})$ is the effective number of degrees of freedom at the temperature of reheating.

Therefore, in those models considered here that specify the reheating history (Cases B \& C), once the model parameters are fixed, $N_*$ is fixed as well (modulo some variation due to uncertainty in the energy content of the universe after reheating). In this sense, the system is overconstrained because $\phi_*$ is not truly a free parameter. However, in principle, one could readily use the remaining freedom to tune the shape of the inflaton potential (beyond the quadratic form) to match $N_*$. To keep our treatment as straightforward as possible, instead of including this additional freedom explicitly and applying the mode matching to $N_*$, we simply do neither. Since, again, our main focus is the properties of the spectator field and $\fnl$, this minimally affects our results. In particular, $\fnl$ is rather insensitive to these choices.
We will briefly consider a more general setup, with varying power law index of the inflaton potential, in Section \ref{subsec:obspros}.

\subsection{Future Galaxy Clustering Constraints}

Our MCMC runs exclusively include current CMB constraints.
However, the motivation of this paper is to quantify the constraining power of next-generation measurements in the resulting space of multifield/spectator models allowed by current data. In particular, we are motivated by upcoming galaxy surveys, which, using scale-dependent halo bias, target order unity precision on local primordial non-Gaussianity, $\sigma(\fnl) \lesssim 1$.
Instead of modeling any specific survey, we will simply compare the posterior parameter and observable distributions from Planck data to this approximate level of constraint, $\Delta \fnl \sim 1$.

\section{Results}
\label{sec:results}

\subsection{The Quadratic-Axion in the Horizon Crossing Approximation}

We first consider the quadratic-axion model using the (improved) Horizon Crossing Approximation, see Section \ref{subsec:intro case A} and Appendix \ref{app:HCA}.
The predictions of $\fnl$ for this model will turn out to be very sensitive to the upper bound chosen for the ``axion decay constant'' $f$. Since in a UV complete theory, it may be difficult to
generate axion-like potentials with $f$ larger than the Planck scale \cite{Banks:2003sx,Svrcek:2006yi}, our default choice will be $f < M_{\rm pl} = \sqrt{8 \pi} \, \mpl$ (note that $M_{\rm pl}$ is not the reduced Planck mass here).
To illustrate the dependence on this cutoff, we will also show results for the prior $f < 3 M_{\rm pl}$ (see Appendix \ref{subsec:app case A} for the other parameter priors).

\begin{figure}[h!]
\centering
\includegraphics[width=.4\textwidth]{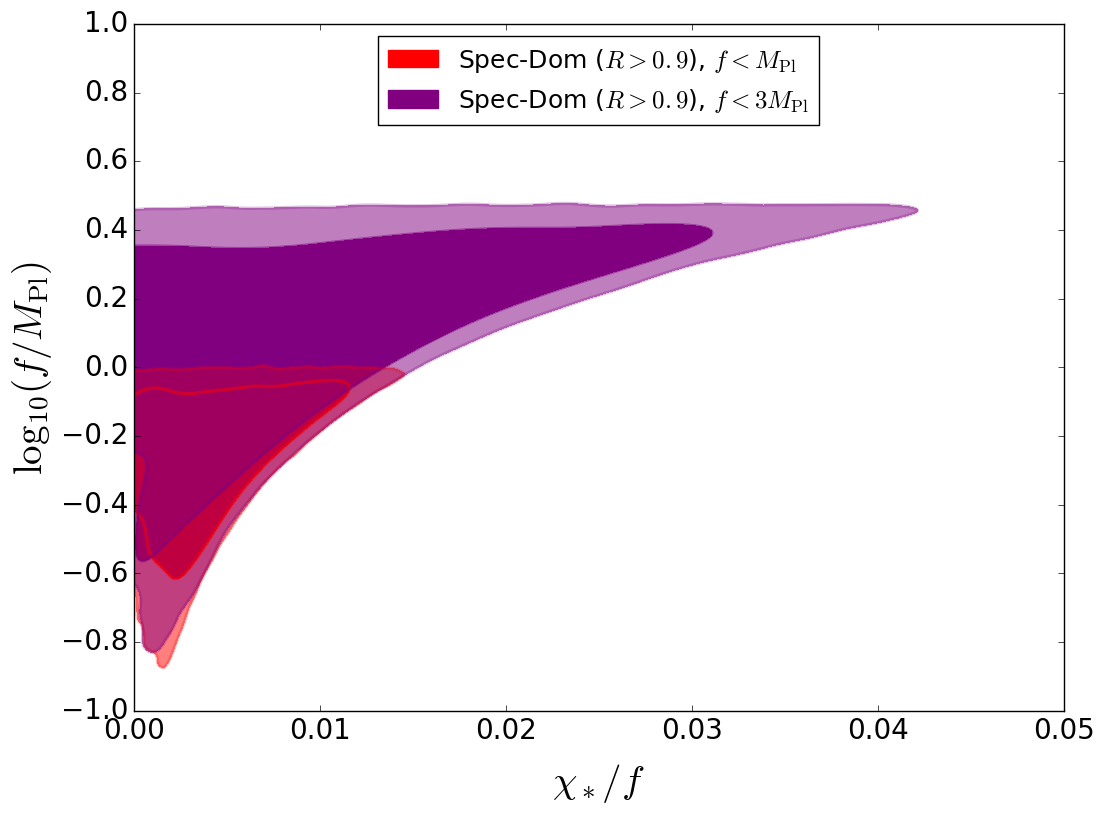}
\caption{The parameter region where the spectator field dominates the curvature perturbations ($R > 0.9$) in the quadratic-axion/HCA model (Case A). Unless otherwise stated, we show $68\%$ and $95\%$ C.L.~contours. We show results for two choices of prior, namely an upper bound on the ``axion decay constant'' $f<M_{\rm Pl}$ in red,
and in purple a bound $f<3 M_{\rm Pl}$. The requirement $\chi_*/f \ll f$, cf.~Eq.~\eqref{chisR} and discussion in main text, is apparent.}
\label{fig:chisvsf}
\end{figure}

In order to gain insight on what the allowed parameter space looks like, let us highlight what imposing the spectator-dominated regime means. Within the HCA, $R$ has a very simple form and $R>0.9$ translates into,
\be
\frac{\chi_*}{f} < \frac{f}{3\pi^2\phi_*}\,.
\label{chisR}
\ee
This behavior is clearly visible (specifically the
contour edges at bottom-right) from Fig.~\ref{fig:chisvsf}, where we plotted the 2D $68\%$ and $95\%$ confidence level (C.L.) contours from our MCMC chains in the plane $(\chi_*/f,f)$, in the spectator-dominated (Spec-Dom) regime.
Note that because of the form of the expressions in Eq.~\eqref{eq:NHCA}, $R$ is independent\footnote{In the HCA, the only place where $V_0$ explicitly appears is in the spectral index $n_s$, see Eq.~\eqref{eq:ns}.} of $V_0$.
The upper bounds in the vertical direction in Fig.~\ref{fig:chisvsf} come directly from the priors on $f$.
Throughout this paper, since bounds in the posterior parameter space are partially determined by (broad) priors, not just by the Planck measurements, the shapes of the posterior distributions commonly deviate from the narrow, Gaussian distributions one may find in a completely data dominated case with small error bars.

Since we envision $\chi/f$ as an axion {\it phase}, our prior expectation is for $\chi_*$ to be uniformly distributed in the interval $[0,f/2]$. Therefore, the requirement of very small $\chi_*/f$ in Eq.~\eqref{chisR} corresponds to significant fine-tuning of initial conditions. Indeed, if we do not explicitly impose $R > 0.9$ in our MCMC runs, this condition is satisfied less than $1\%$ of the time, while most of the points in the chains are concentrated in the inflaton-dominated $R\ll1$ region, with $\fnl \approx 0$. However, this region corresponds precisely to the case that is very similar to single-field inflation. In order to explore the features that are specific to the presence of the extra field, we will now focus on the spectator-dominated regime ($R > 0.9$), keeping in mind that this is a fine-tuned subset of models. In this regime, we are pushed towards large values of $f$, close to the prior upper bound, because large $f$ allows for a larger range of initial field values satisfying Eq.~\eqref{chisR}.

The phenomenology of the background energy densities in the spectator-dominated parameter regime of $f \lesssim M_{\rm pl}$, $\chi_*/f \lesssim f/(3 \pi^2 \phi_*)$ depends on the value of $V_0/m_\phi^2$. For the largest values of this quantity allowed by the spectator requirement and by the constraint on $\eta^\chi_*$ coming from $n_s$, $\rho_\chi$ starts decaying like $a^{-3}$ slightly before the end of inflation.
However, given our broad prior, most of the posterior volume corresponds to much smaller values of $V_0/m_\phi^2$. In that regime, $\chi$ starts oscillating well after inflation, when $\phi$ is already oscillating itself and $\rho_\phi$ decays like matter. In particular, for $f = M_{\rm pl}$ and $\chi_*/f = f/(3 \pi^2 \phi_*)$, $\chi$ comes to dominate the total energy density of the universe before it starts oscillating, leading to a short second phase of inflation (the second scenario in Figure \ref{fig:QAdens}). Lowering $f$ (still with $\chi_*/f = f/(3 \pi^2 \phi_*)$ and  still assuming the low $V_0/m_\phi^2$ regime), the ratio $\rho_\chi/\rho_\phi$ at the time when $\chi$ starts oscillating goes down, and there is no second phase of inflation once $f \lesssim 0.1 M_{\rm pl}$.

\begin{figure}[h!]
\centering
\includegraphics[width=.49\textwidth]{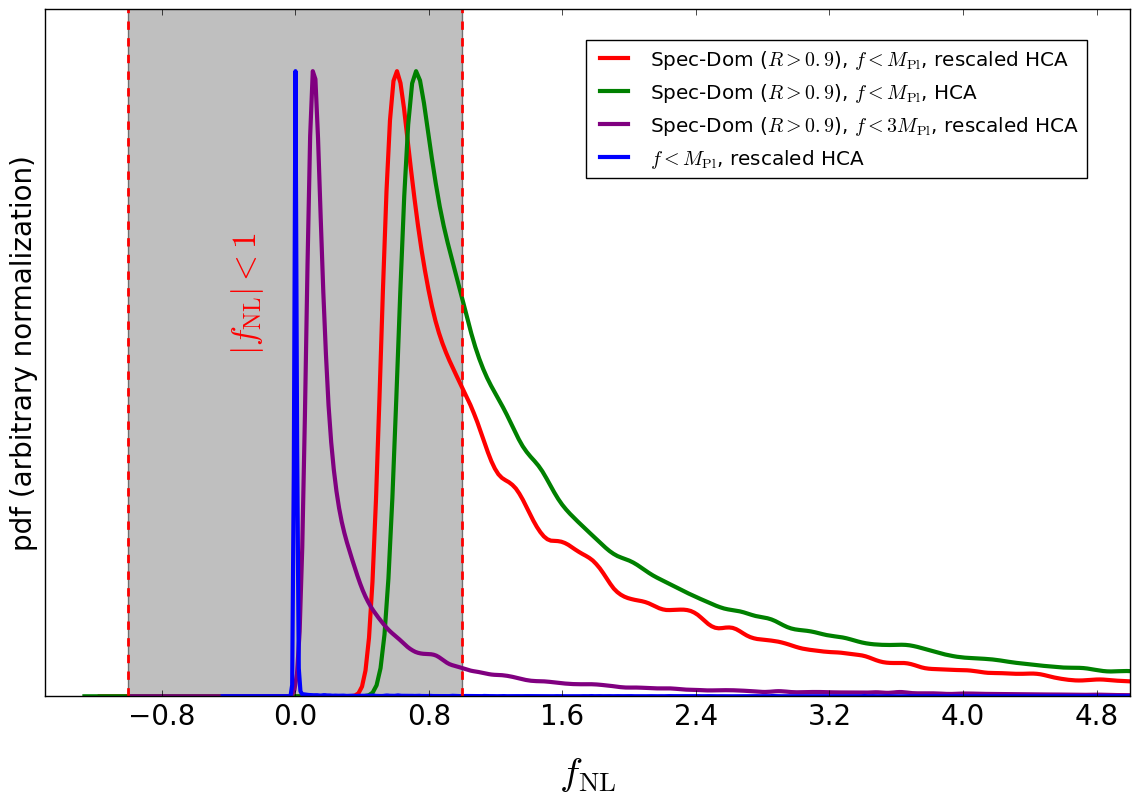}
\caption{Posterior distribution of $\fnl$ in quadratic-axion scenario (Case A), assuming Planck constraints on $\fnl$, $n_s$ and $r$. The blue curve shows the general $\fnl$ distribution in this model, dominated by the inflaton-dominated regime, where $\fnl \approx 0$.  In red and green (see main text and Appendix for discussion of the two approaches), we impose the condition that the curvature perturbations are dominated by the spectator $\chi$ ($R > 0.9$). In this regime, while the exact shape of the $\fnl$ posterior is prior dependent (cf.~purple curve), $\fnl$ is generically of order unity (58\% probability of $|\fnl| > 1$ for our default prior $f < M_{\rm pl}$, red curve).}
\label{fig:fnlQuadAxion_simple}
\end{figure}

In Fig.~\ref{fig:fnlQuadAxion_simple} we plot the posteriors for $\fnl$
for the two different choices of upper bound on $f$.
As discussed in Appendix \ref{app:HCA}, we show results both assuming the standard Horizon Crossing Approximation,
and the improved approximation calibrated on numerical calculations.
The curves for the default prior $f< M_{\rm Pl}$ are well within the Planck CMB bound,
with typical values of $\fnl$ of order unity. This is thus within range of future experiments, especially if $\sigma(\fnl) $ could be pushed significantly below one.

We can understand the $\fnl$ distribution better by noting
that, in the fully spectator-dominated limit, cf.~Eq.~\eqref{eq:fnlHCA},
\beq
\fnl \approx  \frac{5 \pi^2}{3 f^2}.
\eeq
The maximum value $f = M_{\rm pl}$ then corresponds to $\fnl = -0.65$, thus explaining the cutoff in the
$\fnl$ distribution. This cutoff is smoothed out because we consider the range $R = 0.9 - 1$ and the expression for $\fnl$ above is to be multiplied by $R^2$. The relation between $f$ and $\fnl$ also makes clear that the posterior is dominated by a limited range of $f$ just below and up to the cutoff.
Therefore, it is the prior of a sub-Planckian decay constant that pushes us towards $|\fnl| \gtrsim 1$ (assuming the perturbations are dominated by the spectator field in the first place).
For the more inclusive prior, $f < 3 M_{\rm pl}$, the typical value of $\fnl$ is significantly smaller.
The low non-Gaussianity at large $f$ can be understood by noting that in this limit, $V(\chi)$ becomes more and more like a flat, slow-roll potential, which naturally has slow-roll suppressed $\fnl$.

\begin{figure}[h!]
\centering
\includegraphics[width=.49\textwidth]{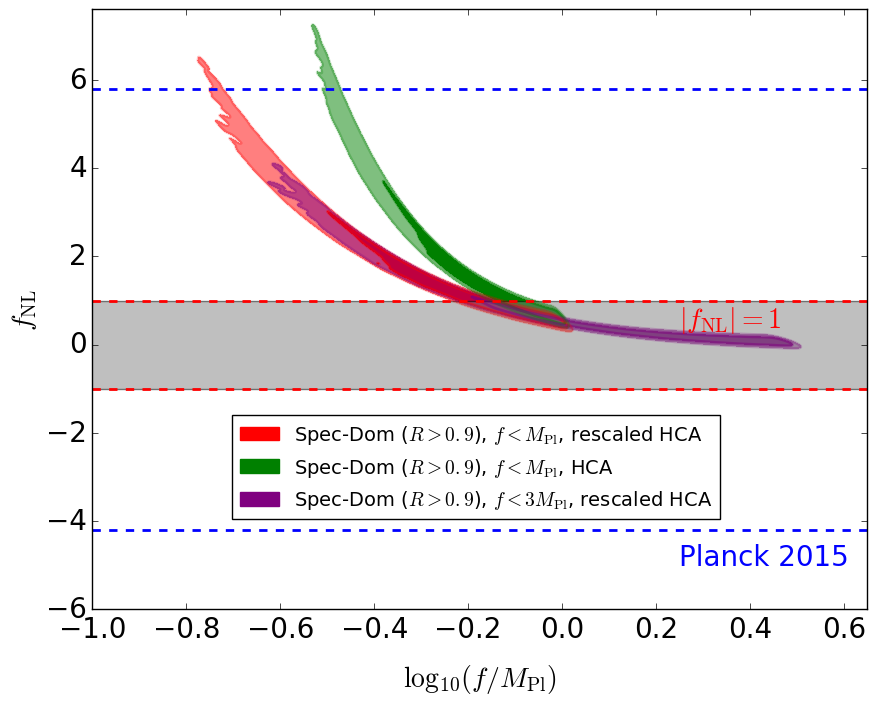}
\caption{Joint posterior distribution of $\fnl$ and ``axion decay constant'' $f$ in quadratic-axion model. The red (purple) contours show the posterior restricted to the spectator-dominated regime with a prior $f<M_{\rm Pl}$ ($f < 3\, M_{\rm Pl}$). The green contour shows the same, but without the correction factor applied to the HCA prediction.
The dashed horizontal lines indicate the current $1\sigma$ range from Planck, and the level of the constraints aimed for by future galaxy surveys, $|\fnl| \sim 1$.}
\label{fig:QuadAxion fnlf}
\end{figure}

We illustrate the relation between $f$ and $\fnl$ in Figure \ref{fig:QuadAxion fnlf}, which shows the joint posterior distribution of $f$ and $\fnl$ in the spectator-dominated regime.
This figure also clearly illustrates
the difference in the dependence of $\fnl$ on $f$ between the HCA and the rescaled/improved HCA.
Note however that the main qualitative conclusions are not strongly affected by whether or not the correction factor is applied and are thus not sensitive to the exact details of the approximation used to compute $\fnl$.

\vskip.1cm

\subsection{The Curvaton}

\begin{figure}[h!]
\centering
\includegraphics[width=.49\textwidth]{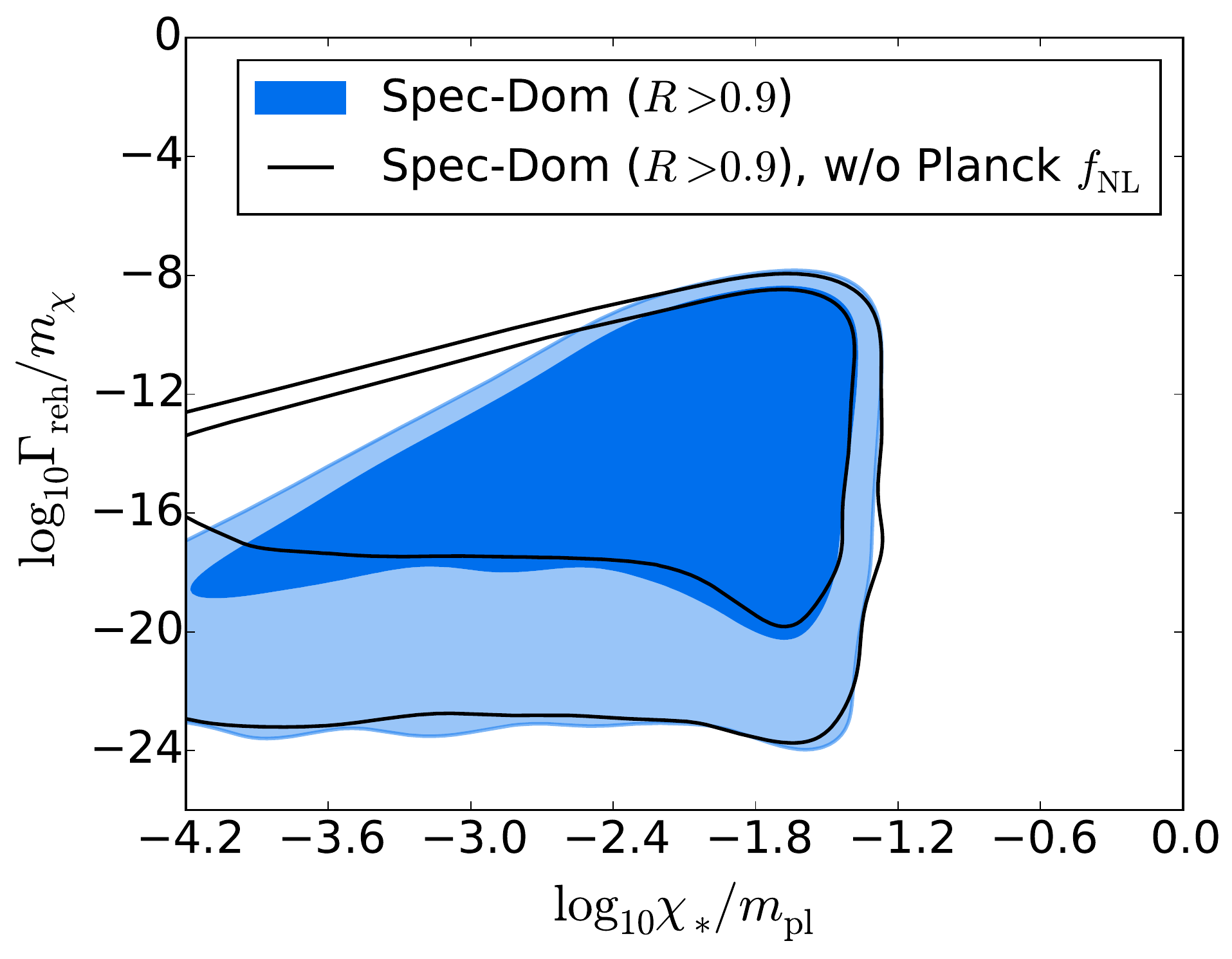}
\caption{Posterior parameter distribution in the spectator-dominated regime of the curvaton model (Case B). The black contours show the constraints without the Planck bound on $\fnl$, while the blue filled contours show the result with all Planck constraints. For the spectator (i.e.~curvaton) to dominate,
reheating needs to occur very long after the start of oscillations in the curvaton (extremely low $\Gamma_{\rm reh}/m_\chi$), representing a fine-tuning. The upper bound on $\Gamma_{\rm reh}/m_\chi$ becomes stricter for small initial field values, $\chi_*$, as explained in the text.}
\label{fig:R curv}
\end{figure}

For the curvaton, we find that the spectator-dominated regime is reached for low initial field values $\chi_*$ and low ratios $\Gamma_{\rm reh}/m_\chi$, the latter corresponding to a long reheating phase. This is illustrated in Figure \ref{fig:R curv}, which shows the posterior probability distributions (68 and 95\% confidence level) in the
spectator-dominated regime ($R > 0.9$).
The blue regions include all CMB measurements discussed above, while the unfilled contours are derived without including the Planck $\fnl$ measurement.

The spectator-dominated parameter region can be understood as follows.
The requirement $N_\chi \gg N_\phi$ can be phrased as (cf.~Eqs.~\eqref{eq:Nps spec}, \eqref{eq:Ncs curv}),
\beq
\label{eq:RCurv}
\frac{r_{\chi,{\rm reh}}}{\chi_*} \gg \frac{\phi_*}{\mpl^2}.
\eeq
Since by definition $r_{\chi,{\rm reh}} \leq 1$, we clearly at least
need $\chi_* \, \phi_* \ll \mpl^2$. Assuming this is satisfied, there is in addition the requirement that $r_{\chi,{\rm reh}}$ is not too small. As long as $r_{\chi,{\rm reh}} \ll 1$, it is easy to show that $r_{\chi,{\rm reh}} \sim \chi_*^2 \sqrt{m_\chi/\Gamma_{\rm reh}}/\mpl^2$, translating Eq.~\eqref{eq:RCurv} into the requirement,
\beq
\label{eq:spec req}
\frac{\Gamma_{\rm reh}}{m_\chi} \ll \left( \frac{\chi_*}{\phi_*}\right)^2.
\eeq
Thus, the smaller the value of $\chi_*$, the more the ratio $\Gamma_{\rm reh}/m_\chi$ has to be tuned to extremely small values. In physical terms, we are forced towards low initial field values, but the smaller $\chi_*$ is, the smaller the ratio of curvaton to radiation energy density is at the start of the curvaton phase, and thus the longer the curvaton phase needs to last
 to make the curvaton fraction $r_{\chi}$ non-negligible.

The above explains well the unfilled contours in Figure \ref{fig:R curv}. The blue regions show that when the Planck $\fnl$ bound is added, an additional part of parameter space is excluded. Namely, in the curvaton-dominated regime, and for small $r_{\chi,{\rm reh}}$, we have $\fnl \sim r_{\chi,{\rm reh}}^{-1}$ so that the Planck bound forces
\beq
\frac{\Gamma_{\rm reh}}{m_\chi} \lesssim \left(f_{\rm NL, max}^{\rm Planck}\right)^2  \, \chi_*^4
\eeq
(where the $2\sigma$ Planck bound is $f_{\rm NL, max}^{\rm Planck} \sim 10$), thus explaining the steeper scaling of the maximum value of
$\Gamma_{\rm reh}/m_\chi$ with $\chi_*$ in the filled blue regions.

{\it Is the spectator-dominated regime fine-tuned?} We have seen above that to satisfy the condition of large $R$, one needs an extremely large hierarchy between the scales $m_\chi$ and $\Gamma_{\rm reh}$, translating to a reheating scale many orders of magnitude below the inflation scale. In this sense, the regime where the spectator/curvaton is important is very fine-tuned. Moreover, we require small initial field values in Planck units. At the same time, we find the posterior probability for, say, $R > 0.5$ vs.~$R < 0.5$, to be of the same order\footnote{The probability of being in the spectator-dominated regime is increased somewhat by the upper bound on the tensor-to-scalar ratio, but this is partially an artifact of our choice of a quadratic inflaton potential, which in the inflaton-dominated regime is in tension with the data (one can fit $n_s$ at the cost of too large a value of $r$).
We have considered the more general case of a varying inflaton potential power law index, see Section \ref{subsec:obspros}, and find that in this case, the probability of being in the spectator-dominated regime is somewhat suppressed compared to the $\phi^2$ model.}. The reason for this is that we imposed logarithmic priors on $\Gamma_{\rm reh}$, etc, with very small lower bounds, reflecting the huge hierarchy between the minimum allowed reheating scale (here chosen to be $H_{\rm reh} \sim 10^{-13}$ GeV, corresponding to $T_{\rm reh} \sim 1$ TeV, see Appendix \ref{app:priors}) and the Hubble scale at the end of inflation, $H_{\rm end} \sim 10^{13}$ GeV.
We also find that the posterior distribution of $R$ (not shown) is bimodal, with peaks at $R = 0$ and $R = 1$. This is again a prior driven effect. There is simply a large parameter volume in the regions where either the spectator or inflaton domination conditions are saturated, cf.~e.g.~Eq.~(\ref{eq:spec req}), and only an order of magnitude of parameter range in the intermediate regime.

\begin{figure}[h!]
\centering
\includegraphics[width=.49\textwidth]{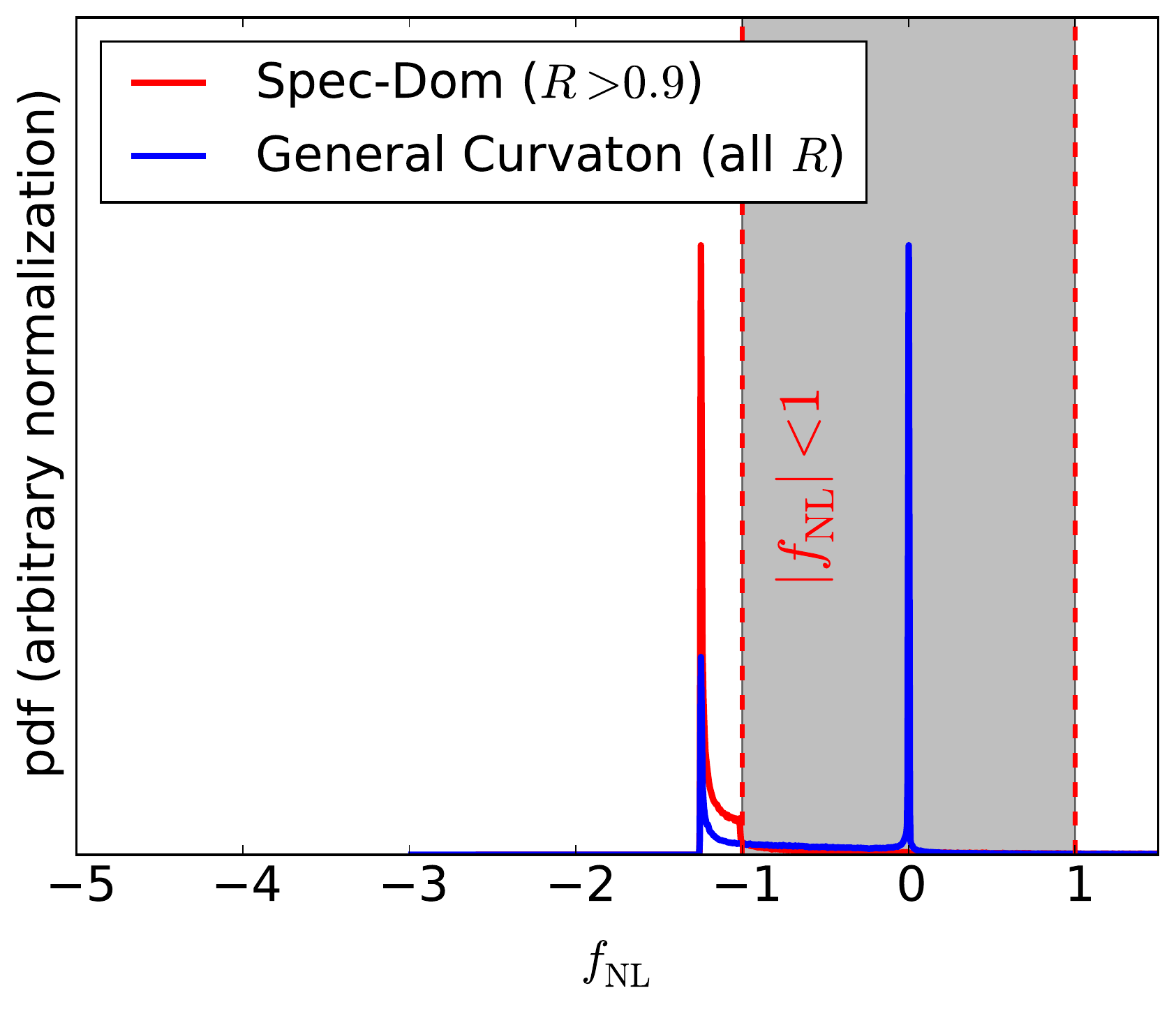}
\caption{Posterior distribution on $\fnl$ in curvaton scenario (Case B), assuming Planck measurements of $\fnl$, $n_s$ and $r$. The blue curve shows the general $\fnl$ distribution in this model, while the red curve is restricted to spectator domination, where curvature perturbations are mostly sourced ($R > 0.9$) by the curvaton $\chi$. In this regime, wile the exact shape of the $\fnl$ posterior is sensitive to priors, $\fnl$ is generically of order unity (79 \% probability of $|\fnl| > 1$). In particular, most of the posterior probability is in the parameter region where the curvaton phase has lasted long enough for the curvaton to dominate the background energy density of the universe, $r_{\chi, {\rm reh}} \to 1$, so that the non-Gaussianity reaches its asymptotic level, $\fnl \to -5/4$.}
\label{fig:fnlcurvaton_simple}
\end{figure}

In Figure \ref{fig:fnlcurvaton_simple}, we consider the posterior distribution of $\fnl$ both in the general model,
and in the spectator dominated regime.
In the latter case (red), the peak corresponds to the scenario where the curvaton stage lasts long enough for the curvaton to dominate the energy budget of the universe, $r_{\chi,{\rm reh}} \to 1$ and $\fnl \to -5/4$. We find that 79\% of the posterior distribution has $|\fnl| > 1$, making future constraints at this level extremely interesting. In particular, $\fnl = -5/4$ is clearly an important target.

The dominance of the peak at $\fnl = -5/4$ reflects that our priors allow a large parameter volume where $r_{\chi,{\rm reh}} = 1$ is saturated, i.e.~once $\Gamma_{\rm reh}$ is low enough for the curvaton to dominate the background energy, lowering $\Gamma_{\rm reh}$ further by orders of magnitude will maintain $\fnl = -5/4$.
If we had imposed priors that penalize a large hierarchy between $\Gamma_{\rm reh}$ and $m_\chi$, the results would change, favoring the large negative $\fnl$ regime (low $r_{\chi, {\rm reh}}$) relative to $\fnl = -5/4$. Of course, such a change in priors would also make satisfying the condition of large $R$ more manifestly fine-tuned.
In the general case (blue curve), we see the aforementioned bimodality of the posterior of $R$, with the inflaton-dominated regime leading to the single-field value $\fnl \approx 0$ and the spectator-dominated case giving $\fnl = -5/4$.

\begin{figure}[h!]
\centering
\includegraphics[width=.49\textwidth]{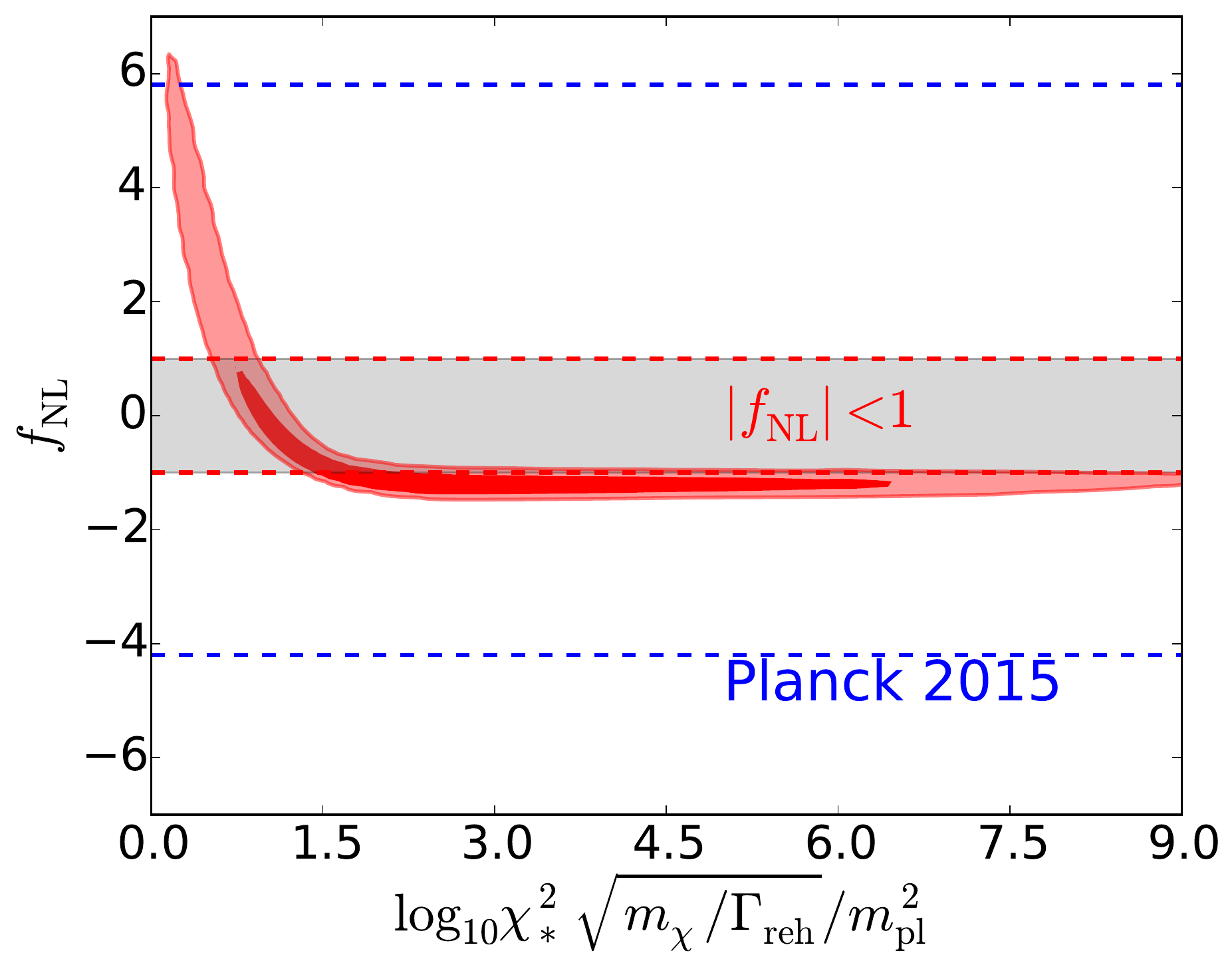}
\caption{Posterior distribution of $\fnl$ vs.~combination of model parameters in the curvaton model, restricted to the spectator-dominated regime ($R > 0.9$).
The parameter combination $\chi_*^2\sqrt{m_\chi/\Gamma_{\rm reh}}/\mpl^2$ is approximately equal to the ratio $r_{\chi,{\rm reh}}$ for low $r_{\chi,{\rm reh}}$ (whereas large values of $\chi_*^2\sqrt{m_\chi/\Gamma_{\rm reh}}/\mpl^2$ correspond to $r_{\chi,{\rm reh}} = 1$), making it a good proxy for $\fnl$.
The dashed horizontal lines indicate the current $1\sigma$ limits from Planck, and the constraints aimed for by future galaxy surveys, $|\fnl| \sim 1$.}
\label{fig:curv absfnl}
\end{figure}

Now focusing on the spectator-dominated regime, Figure \ref{fig:curv absfnl} shows the joint posterior distribution of $\fnl$ with the parameter combination $\chi_*^2 \sqrt{m_\chi/\Gamma_{\rm reh}}/\mpl^2$, which, as explained above, is approximately equal to $r_{\chi, {\rm reh}}$ for low values of $r_{\chi, {\rm reh}}$.
A future measurement of $\fnl$ with order unity precision thus may provide important information on the curvaton model, and in particular on this parameter combination.

\subsection{Modulated Reheating}
\label{subsec:MR results}

\begin{figure}[h!]
\centering
\includegraphics[width=.49\textwidth]{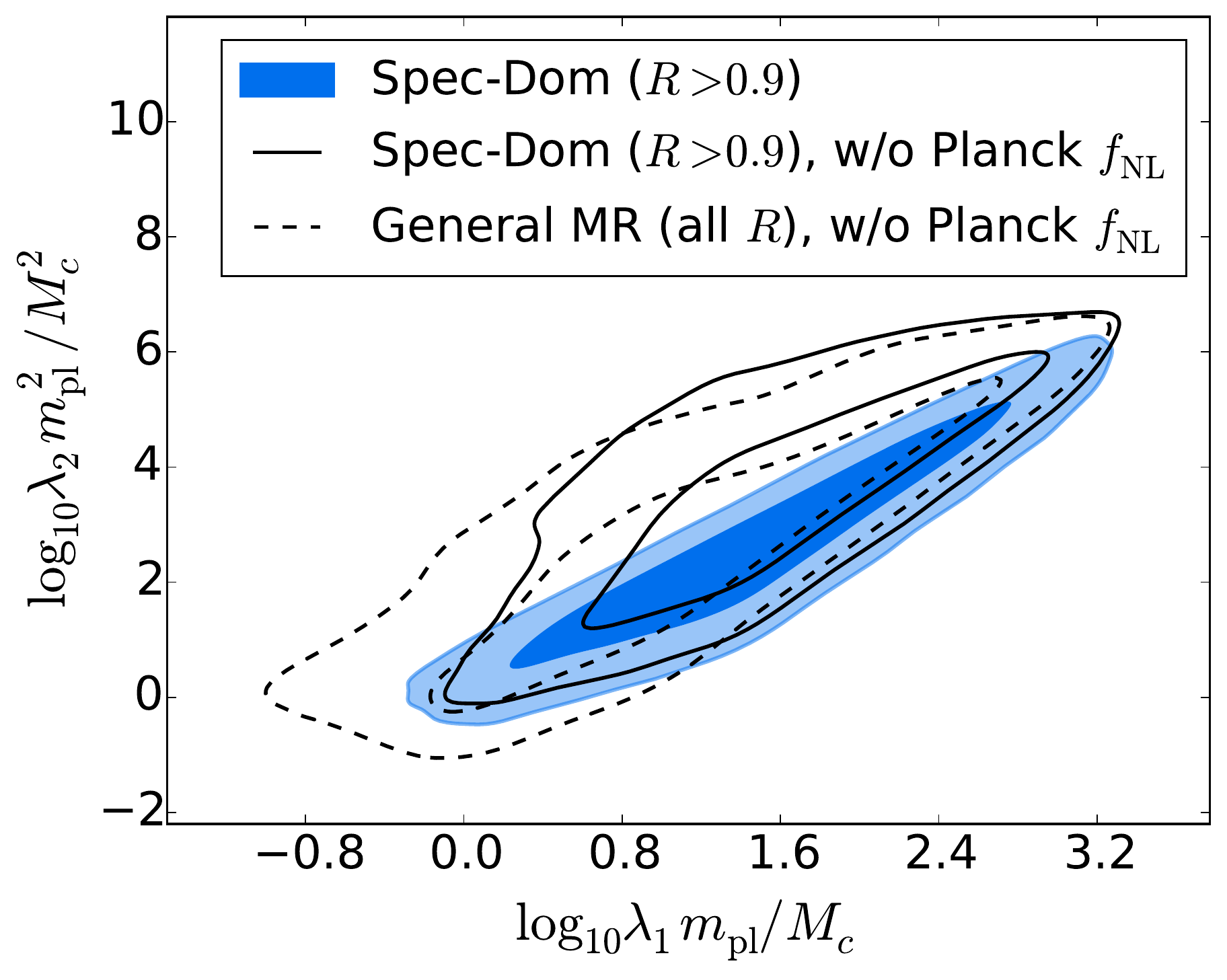}
\caption{Spectator-dominated regime ($R > 0.9$) of the modulated reheating model (Case C).
The contours show the posterior distribution of the first- and second-order coefficients of $\chi$ in the inflaton coupling constant determining the reheating decay rate, see Eq.~(\ref{eq:lambda def}).
The black dashed contours depict constraints without the Planck bound on $\fnl$ for general $R$.
The strong correlation between the two parameters shown is due to the joint dependence on the cutoff mass $M_c$.
The black solid contours additionally require $R > 0.9$.
This spectator-dominated regime thus corresponds to strong dependence of $\lambda$ on $\chi$.
The filled blue contours finally add the Planck $\fnl$ limit, limiting the final allowed region to lower values of $\lambda_2/M_c^2$.}
\label{fig:R MR}
\end{figure}

In the modulated reheating model, the spectator-dominated regime is reached if (cf.~Eq.~(\ref{eq:Ncs MR})),
\beq
\left| \frac{\lambda'}{\lambda} \right| \gg \frac{\phi_*}{\mpl^2}
\eeq
(the transfer function from $\chi_*$ to $\chi_{\rm reh}$ generally has a small effect),
corresponding to large $\lambda_1/M_c$ and/or $\lambda_2 \chi_{\rm reh}/M_c^2$ (most of the weight in the prior distribution of $\lambda_0$ lies around values of $\mathcal{O}(10^{-1})$ because of the uniform prior). This region is shown in Figure \ref{fig:R MR}. The filled contours show the usual confidence regions with the prior $R > 0.9$, including all Planck data discussed, while the solid empty contours represent the same region, but without the $\fnl$ bound.

We have also (dashed empty contours) included the posterior in the general model, i.e.~without the spectator domination requirement on $R$ (and also without the $\fnl$ measurement included), to illustrate that the strong correlation between $\lambda_1/M_c$ and $\lambda_2/M_c^2$ is there regardless of the requirement on $R$. It is mostly prior driven, and comes from the fact that both quantities scale with the same cutoff mass $M_c$ (and that the dimensionless quantities $\lambda_1$ and $\lambda_2$ follow uniform priors). What the requirement of spectator domination does is to shift $\lambda_1/M_c$ and $\lambda_2/M_c^2$ to larger values along the correlation direction, as shown by the solid black contours and filled blue regions.

Thus, spectator domination requires an effective cutoff scale $M_c$ not much larger than $\sim 0.1 \, \mpl$, allowing the effect of $\chi$ on $\lambda$ to be large enough. Since we do not want any contribution to $\lambda$ to be larger than unity, the requirement of large $\lambda_1/M_c$ and $\lambda_2/M_c^2$ in Planck units does again mean we need small initial field values, $\chi_* \ll \mpl$, which can be considered fine-tuning. For the same reasons discussed in the curvaton case, related to our choice of priors, our chains do give a bimodal distribution of $R$ with peaks of comparable amplitude at $R=0$ and $R=1$ despite this fine-tuning.

\begin{figure}[h!]
\centering
\includegraphics[width=.49\textwidth]{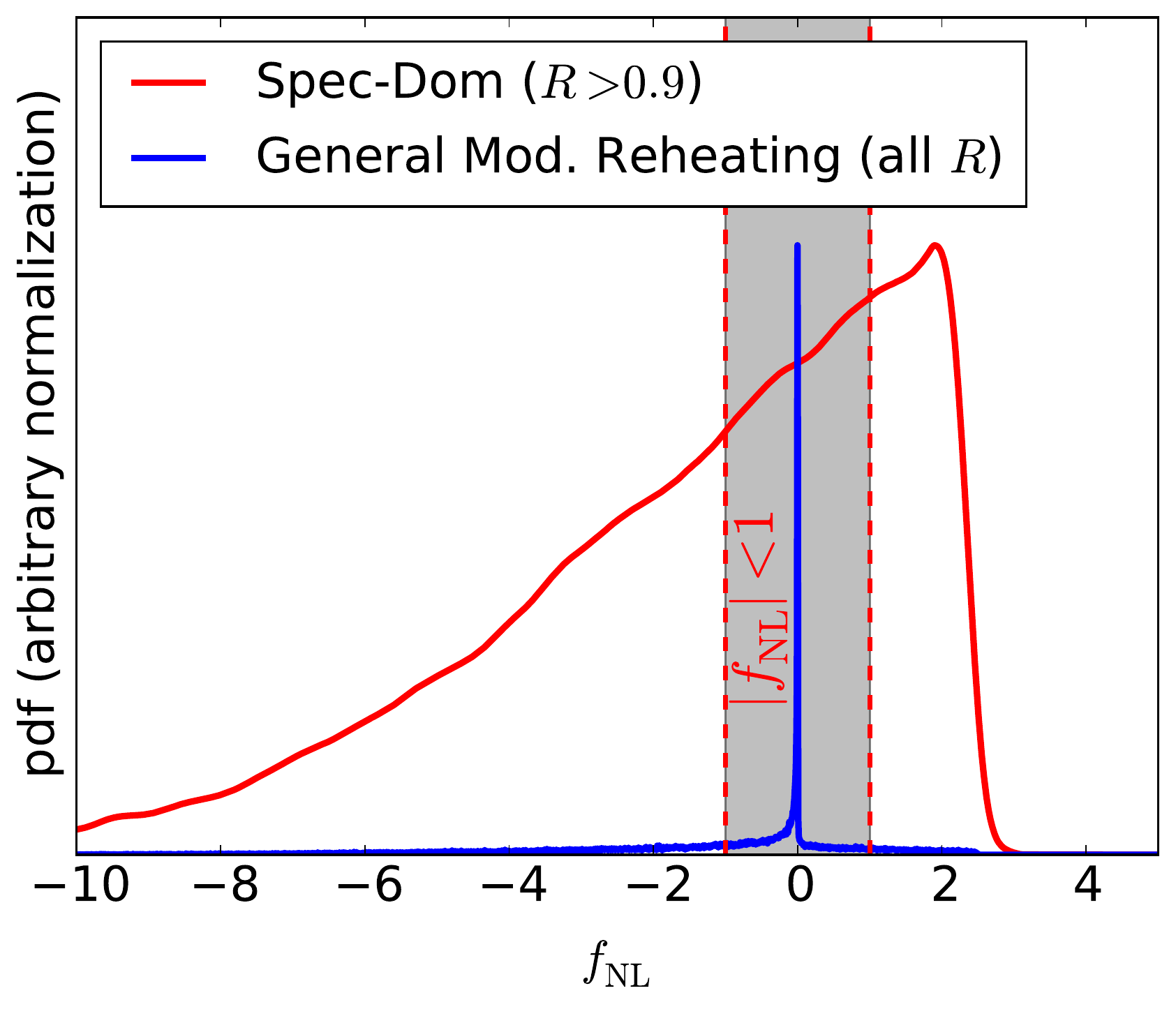}
\caption{Posterior distribution of $\fnl$ in modulated reheating model (Case C), assuming Planck constraints on $\fnl$, $n_s$ and $r$. The blue curve shows the general $\fnl$ distribution in the model, while the red curve is restricted to spectator domination ($R > 0.9$), where fluctuations in $\chi$ dominate the final curvature perturbations due to their effect on reheating. In the former (general) case, a large fraction of the posterior probablity lies in the inflaton-dominated regime with negligible $\fnl$. In the latter case,
while the exact shape of the $\fnl$ posterior is sensitive to priors, $\fnl$ is generically of order unity (72\% posterior probability of $|\fnl| > 1$), with a broad distribution.
}
\label{fig:fnl_modulated_reheating}
\end{figure}

Figure \ref{fig:fnl_modulated_reheating} depicts the posterior distribution of $\fnl$ for both the general case and the spectator-dominated case. In the latter case, we see a relatively broad distribution of values (in contrast with the curvaton model), with typical values of order $|\fnl| \sim 1 - 5$ (we note that when we do not implement the current observational bound on $\fnl$, the distribution is significantly broader (not shown), with typical values of order $|\fnl| \sim 10 - 20$). We find that 72\% of the parameter space in the spectator-dominated regime has $|\fnl| > 1$.

The distribution at the lower end has a relatively sharp cutoff. This follows from the specific form of the expression for $\fnl$, Eq.~(\ref{eq:fnl MR}). Ignoring the evolution of $\chi$, it reduces to
\beq
\label{eq:fnl MR simple}
\fnl \approx 5 \left( 1 - \frac{\Gamma''_{\rm reh} \, \Gamma_{\rm reh}}{(\Gamma'_{\rm reh})^2 }\right)
= \frac{5}{2} \left( 1 - \frac{\lambda''(\chi_{\rm reh}) \, \lambda(\chi_{\rm reh})}{(\lambda'(\chi_{\rm reh}))^2}\right).
\eeq
Since we have chosen the coefficients in the expansion of the reheating coupling to all be positive, this gives an upper bound $\fnl < 5/2$. This cutoff gets smoothed out once the evolution of $\chi$ is included (the partial derivatives in Eq.~(\ref{eq:fnl MR})), thus explaining the shape of the red curve at the high $\fnl$ end.

In the general case (blue), the bimodal distribution of $R$ again leads to a superposition of the inflaton-dominated regime's $\fnl \approx 0$ and the broader distribution corresponding to the spectator-dominated regime.

\begin{figure}[h!]
\centering
\includegraphics[width=.49\textwidth]{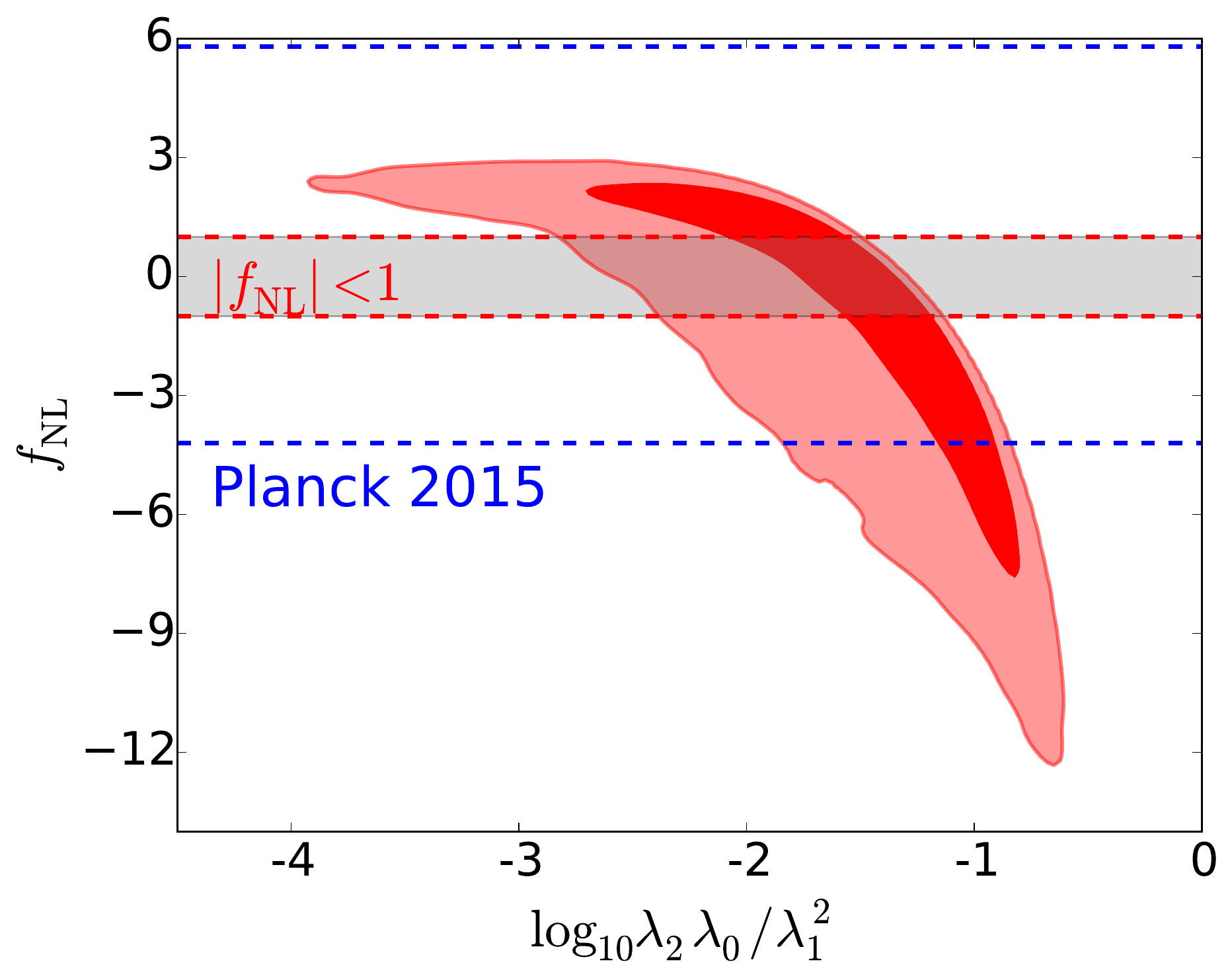}
\caption{Joint posterior distribution of $\fnl$ and a combination of parameters describing the inflaton reheating decay rate and its dependence on the field $\chi$, assuming spectator domination ($R > 0.9$). A measurement of $\fnl$ provides valuable information on the modulated reheating process, and in particular the quantity $\lambda_2 \lambda_0/\lambda_1^2$, cf.~Eq.~(\ref{eq:fnl MR}).
The dashed horizontal lines indicate the current $1\sigma$ limits from Planck, and the constraints aimed for by future galaxy surveys, $|\fnl| \sim 1$.}
\label{fig:MR absfnl}
\end{figure}

Studying $\fnl$ in the spectator-dominated scenario in more detail, Figure \ref{fig:MR absfnl} shows the joint posterior of $\fnl$ and $\lambda_2 \lambda_0/\lambda_1^2$. This parameter combination mostly determines $\fnl$ in the spectator-dominated regime if $\lambda'$ is dominated by the $\lambda_1$ contribution and $\lambda$ by $\lambda_0$, cf.~Eq.~(\ref{eq:fnl MR simple}). A measurement of $\fnl$ provides information on the modulated reheating parameter space and in particular on this combination of parameters describing the coupling of the inflaton to $\chi$ and to the particles into which it reheats.

In summary, while the physics behind the mechanisms is very different, the modulated reheating has similar phenomenology to the curvaton scenario. The main qualitative difference is that for spectator-dominated modulated reheating, the $\fnl$ distribution does not peak at a special value ($\fnl = -5/4$ for the curvaton). Instead, it has a broader distribution, with a ``smooth'' cutoff around $\fnl \sim 5/2$.

\subsection{Observational Prospects}
\label{subsec:obspros}

\begin{figure}[h!]
\centering
\includegraphics[width=.5\textwidth]{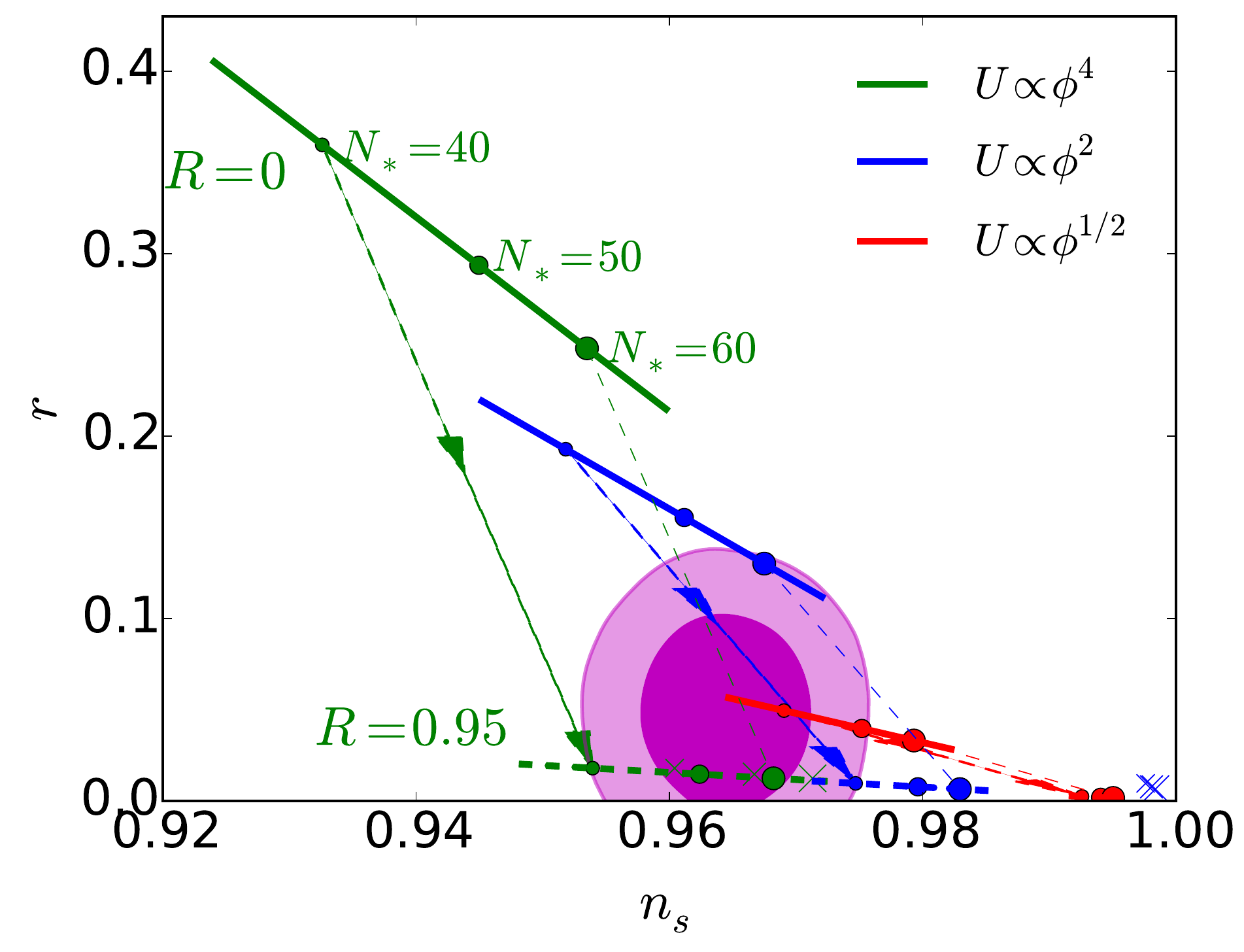}
\caption{Phenomenology in the $(n_s, r)$ plane of models with power-law inflaton potentials, extended beyond the single-field case into the spectator domain ($R > 0$). Magenta contours show current Planck constraints. Solid lines show predictions for inflaton-dominated models ($R=0$, equivalent to single-field). $N_*$ is the number of $e$-folds to the end of inflation. The arrows connect the $R=0$ regime to the spectator-dominated regime (here, $R=0.95$, shown in dashed). This is for the common case of $\eta^\chi_* \approx 0$, while crosses indicate the alternative of large $\eta^\chi_*$. Otherwise ruled out inflaton potentials, such as $U(\phi)\propto \phi^4$ become viable again in the spectator-dominated regime.}
\label{fig:ns vs r}
\end{figure}

Spectator models are a relatively simple extension of singl$e$-field inflation, which itself can be seen as the inflaton-dominated corner of spectator model parameter space.
Regarding the inflaton potential, $U(\phi)$,  we have so far focused on the simple quadratic potential
because predictions for $\fnl$ are rather robust against the details of the inflaton potential.
Technically, however, to fit $n_s$ and $r$ well with realistic values of the number of $e$-folding before the end of inflation,
more freedom in the shape of the inflaton potential is needed. In particular, let us consider the class of power-law models,
\beq
U(\phi) \propto \phi^n,
\eeq
where we will allow non-integer values of $n$.

\begin{figure}[h!]
\centering
\includegraphics[width=.47\textwidth]{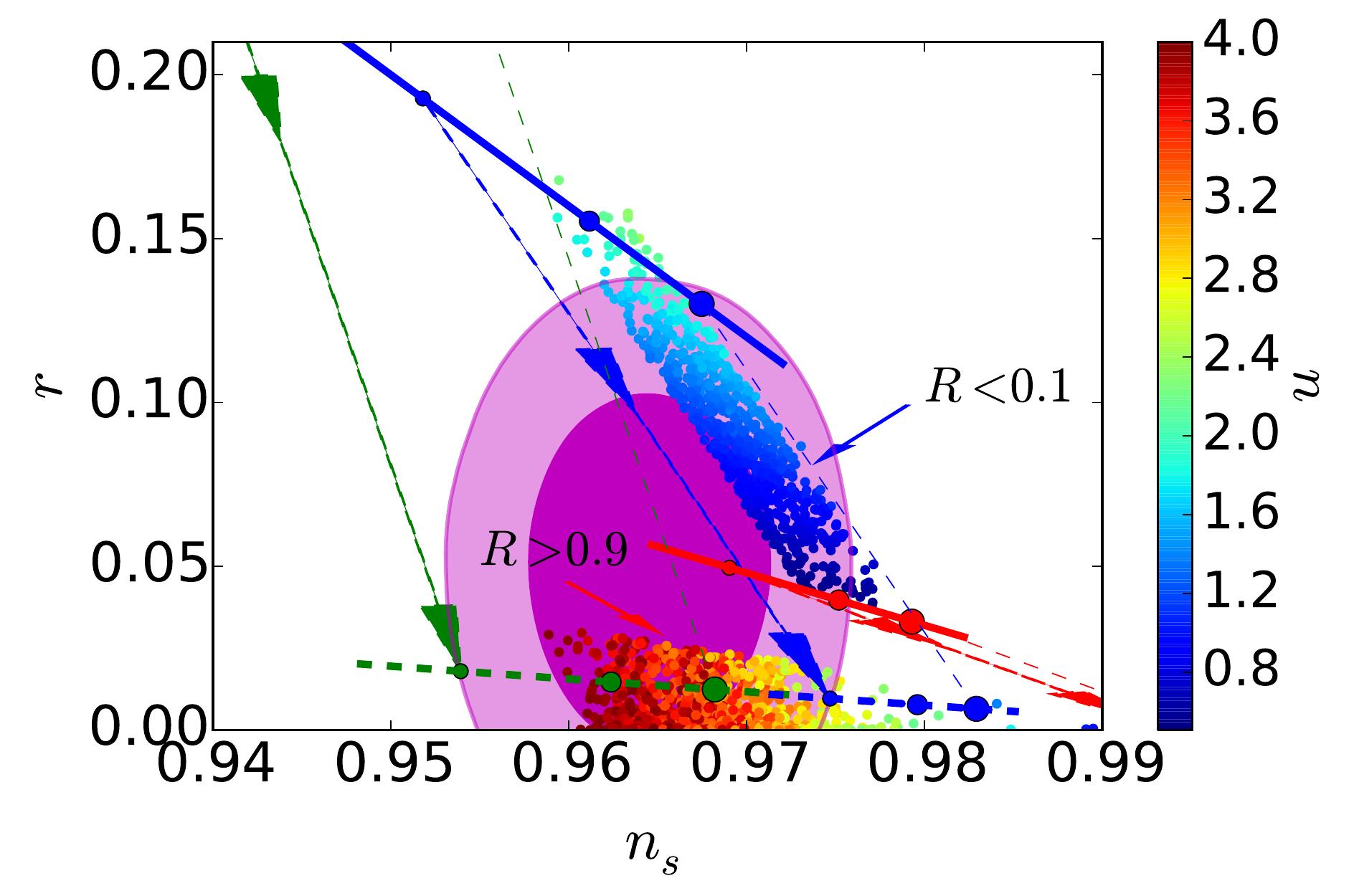} 
\includegraphics[width=.47\textwidth]{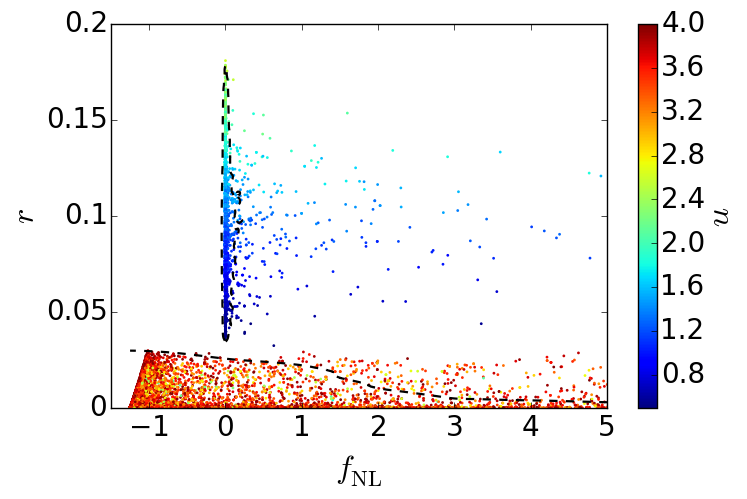}
\caption{{\it Top:} As Figure \ref{fig:ns vs r} (different scale and omitting labels), but with results from MCMC analysis of curvaton model (Case B) added in. Here, the power law index $n$ of the inflaton potential is treated as a free parameter and indicated by color. The two sets of points from the Monte Carlo chains correspond to inflaton domination ($R < 0.1$, blue) and spectator domination ($R > 0.9$, yellow-red).
The former prefer a low index ($n \approx 0.5 - 1$) as in the single-field case, while the latter prefer a large index ($n \approx 3 - 4$). {\it Bottom:} For the same two regimes ($R < 0.1$ and $R > 0.9$), now we show samples with $\fnl$ instead of $n_s$ on the horizontal axis (dashed lines outline the $95 \%$ C.L. regions). While the spectator-dominated regime suppresses $r$, it leads to potentially detectable $\fnl$.}
\label{fig:ns vs r curv}
\end{figure}

Figure \ref{fig:ns vs r} shows the predictions for such models in the $(n_s, r)$ plane, compared to the Planck constraint.
The solid lines show the well known single-field/inflaton-dominated ($R=0$) case, cf.~e.g.~Fig.~12 in \cite{planck2015}. The dots indicate the number of $e$-folds before the end of inflation, $N_*$.
As is well known, the Planck data are already in significant tension with the inflaton-dominated quadratic model, but lower powers, e.g.~$U \propto \phi^{2/3}$ are in reasonable agreement.

The effect of curvature perturbations more and more generated by the spectator field, i.e.~increasing $R$, is indicated by the arrows, leading to the mostly spectator-dominated scenarios ($R = 0.95$) shown in dashed lines,
cf.~Eq.~(\ref{eq:ns}).
Note that for a given $R$ and a given inflaton potential, $n_s$ does not generally have a fixed value because it still depends on $\eta^\chi_*$. However, we find that the regime with negligible $\eta^\chi_*$ contribution often dominates so that we chose $\eta^\chi_* = 0$ in this plot. To indicate the range of effects from non-zero $\eta^\chi$, the crosses show $(n_s, r)$ for the maximum (positive) $\eta^\chi_*$ consistent with the requirement that the spectator field is slowly rolling until after the end of inflation.

Figure \ref{fig:ns vs r} thus visualizes that, as the spectator field becomes more important,
$r$ goes down, making it easier to evade the tensor-mode constraint, and $n_s$ shifts to larger values.
This means that: (1) models that are currently a decent fit in the inflaton-dominated regime (low $n$ power laws) become poor fits in the spectator-dominated case and (2) models with larger power law indices, ruled out by Planck data in the single-field case, become viable again in the spectator-dominated scenario.

We illustrate this for the curvaton model in Figure \ref{fig:ns vs r curv} (top), which shows the same curves, but in a zoomed-in region. Here, we add the results of MCMC simulations, as above but now treating the power law index $n$ as an additional free parameter with $n = [1/2, 4 ]$ and requiring $N_* = [ 46, 58 ]$.
The colored points show the inflaton-dominated posterior region ($R < 0.1$) and the spectator-dominated one ($R > 0.9$). Colors indicate the potential power law index, confirming the picture described above, with the inflaton/spectator-dominated regimes preferring small/large values of $n$.

A major difference between the two regimes is that, while in the inflaton-dominated case, $r$ is always within reach of upcoming B-mode searches (assuming a power law potential), for $R \to 1$, one can obtain $r$ arbitrarily close to zero while perfectly fitting $n_s$, cf.~Eq.~(\ref{eq:r spec}). This is where primordial non-Gaussianity comes in, as illustrated in the bottom panel of Figure \ref{fig:ns vs r curv}.
While in the (fully) inflaton-dominated regime, the single-field consistency condition effectively sets $\fnl$ to zero (the outlying blue points with non-negligible $\fnl$ are explained by their spectator contribution, i.e.~$R \sim 0.1$ and $\Ncscs/\Ncs^2$ very large), the spectator-dominated regime typically generates $|\fnl| \gtrsim 1$.
The same is true for the other two models considered in this paper. The fact that $r$ is typically large in the inflaton-dominated regime is specific to large-field potentials, such as the power laws chosen here. For different types of potentials, it is possible to have small $r$ even in the single-field/inflaton-dominated regime.

In summary, for inflaton-dominated models, $\fnl$ is small and out of reach of near-future experiments, but $r$ is large (assuming a power law inflaton), while in spectator-dominated models, values of $\fnl$ within the scope of upcoming surveys are common, but $r$ is suppressed (we do note that, while in the latter case, large $\fnl$ is expected, it is not impossible to be in the worst-case scenario where both $r$ and $\fnl$ are negligibly small).
Thus, in order to unravel the mysteries of inflation, it is crucial for future probes to aim their sights at both tensor fluctuations and primordial non-Gaussianity.

\section{Discussion \& Conclusions}
\label{sec:concl}

Upcoming galaxy surveys aim to significantly improve constraints on local primordial non-Gaussianity, from the current Planck bound $\fnl = 0.8 \pm 5.0$, to constraints with uncertainties $\sigma(\fnl) \lesssim 1$.
Motivated by this prospect, we have here derived current constraints on a range of multifield inflation models given Planck CMB data and physically motivated parameter priors, and compared the resulting predicted values of $\fnl$ to the expected future constraints.
Our goal was to obtain quantitative estimates, given an inflationary model, of the discovery potential of local non-Gaussianity with these future surveys, and to quantify what such a future $f_{\rm NL}$ may teach us about the physics behind inflation.

We have specifically focused on so-called spectator models, where, while inflation is driven by the inflaton field, the primordial curvature perturbations are partially or fully generated by a second field, the ``spectator''.
At horizon exit, this spectator field does not contribute to the curvature perturbations,
but its perturbations can be converted into curvature perturbations afterward through super-horizon evolution.
We have considered three specific mechanisms for this process with the conversion occurring during different phases: (A) during or after inflation before either field has decayed into radiation, (B) after inflation while the inflaton has already decayed into radiation and the spectator (i.e.~curvaton) oscillates around the minimum of its potential, and (C) after inflation during the reheating process itself.

If the relative contribution of the spectator field to the final primordial curvature power spectrum is close to one, significant non-Gaussianity can be generated, which is why our main focus has been on this set of ``spectator-dominated'' models. While there are significant differences between the three scenarios (A)-(C), we will below discuss some of the main general conclusions.

\vskip.2cm

Typically, to be in the spectator-dominated regime, some form of fine-tuning is required.
For instance, in all three scenarios, small values of the initial spectator field value are required.
Furthermore, in the curvaton scenario, the reheating scale needs to be tuned to be many orders of magnitude below the scale of inflation and the curvaton mass, corresponding to extremely late reheating (although not in clear tension with data).
On the other hand, statements about fine-tuning are always strongly prior dependent. For example, since our MCMC analysis employed wide, logarithmic prior ranges on most dimensionful parameters, we found in both scenarios (B) and (C) that being in the spectator-dominated regime is approximately equally likely as the alternative.
However, this does not remove the objection that large hierarchies between parameters may be unnatural from a model building perspective. Such theory-based prejudice could have been incorporated by modifying our priors, but we chose not to pursue this here.

Assuming spectator domination ($R > 0.9$), we have quantified the posterior distribution of $\fnl$ given current Planck data for each of the three scenarios. We have quantified the promise of next-generation $\fnl$ measurements by quoting the posterior probability of $|\fnl| > 1$, which we will summarize below. Assuming
$|\fnl| > 1$ can be distinguished from zero at sufficient significance, this gives the probability of detection of non-Gaussianity. Conversely, if an upper bound $|\fnl| < 1$ is obtained from the data, the number above tells us what fraction of the currently allowed parameters space will be ruled out.
However, the above quantity does not tell the full story\footnote{Since by default we included the Planck bound on $\fnl$ to compute the posterior, some caution is needed in interpreting the posterior probability of getting $|\fnl| > 1$. If the $\fnl$ posterior {\it without} including the $\fnl$ bound from Planck is very wide compared to $\sigma(\fnl)$ from Planck, the default posterior {\it with} the Planck $\fnl$ bound included is essentially determined by $\sigma(\fnl) \sim 5$, and saying that a bound with $\sigma(\fnl) \lesssim 1$ rules out a large part of currently allowed parameter space is equivalent to the trivial statement that the future error bars are smaller than the current ones. In this scenario, it could be that the Planck $\fnl$ bound had already ruled out an overwhelming fraction of the previously allowed parameter space, and a future tighter bound will simply rule out a little bit more. Therefore, it is also important to quantify how much better a future $\fnl$ constraint does than the current CMB bound.} so below we also quote
what fraction of the posterior distribution obtained {\it without} including the Planck $\fnl$ bound has $|\fnl| > f_{\rm NL, max}^{\rm Planck} = 10$ (corresponding approximately to the $2 \sigma$ Planck bound) and what fraction has $|\fnl| > 1$.\\

\begin{itemize}
\item
{\bf Case A - Quadratic-Axion}\\
With Planck $\fnl$: $P(|\fnl| > 1) = 58 \%$\\ 
Without: $P(|\fnl| > 1 \, (10)) = 63 \,(6)\%$ 
\item
{\bf Case B - Curvaton}\\
With Planck $\fnl$: $P(|\fnl| > 1) = 79 \%$\\ 
Without: $P(|\fnl| > 1 \, (10)) = 83 \,(14)\%$ 
\item
{\bf Case C - Modulated Reheating}\\
With Planck $\fnl$: $P(|\fnl| > 1) = 72 \%$\\ 
Without: $P(|\fnl| > 1 \, (10)) = 92 \,(60)\%$ 
\end{itemize}
We see that in the modulated reheating scenario, the Planck $\fnl$ constraint has already ruled out a significant fraction of the parameter space allowed without taking PNG into account, but that in the other cases we are only just starting to take advantage of $\fnl$.
While, as we have discussed, the numbers above are prior dependent (especially in Case A, which relies on the maximum value of the decay constant $f$ being of order $M_{\rm Pl}$), they suggest that, if inflation is described by one of these models where the curvature perturbations are generated by a field other than the inflaton, future $\fnl$ searches with $\sigma(\fnl) \lesssim 1$ have a good shot at a detection and will probe these models well beyond the current Planck $\fnl$ constraint.

If a detection of $\fnl$ {\it is} achieved, the most important implication would of course be the discovery of multifield inflation (although there are caveats to the single-field consistency conditions that allow non-zero $\fnl$ in certain special single-field scenarios \cite{chenetal13,mooijpalma15}). In addition, we have shown that a measurement of $\fnl$ in the context of the models above also tells us about the values of certain parameter combinations, thus providing hints about the nature of the multifield model describing the early universe.
We have also highlighted the complementarity between B-mode searches constraining primordial tensor perturbations and measurements of galaxies and the CMB constraining primordial non-Gaussianity. It is such a multipronged
approach that provides the best opportunity for improving our understanding of the physics of the extremely early universe.

In conclusion, while large or order unity $\fnl$ is {\it not} a general prediction of multifield inflation, it appears to be quite generic in spectator-dominated models. Arguably, these are the more interesting multifield models regardless of $\fnl$, as the case where the primordial fluctuations are fully determined by the inflaton field is phenomenologically indistinguishable from single-field models.
The appearance of order unity $\fnl$ in spectator-dominated and similar models has been highlighted many times in the literature, but here we have sampled the full parameter space of a range of models, taking into account observational constraints, leading to a more quantitative assessment of the typical prediction for $\fnl$.

\vskip.1cm
\emph{Acknowledgements:}
We would like to thanks P.~Bull and D.~Green for helpful discussions.
Part of the research described in this paper was carried out at the Jet Propulsion Laboratory, California Institute of Technology, under a contract with the National Aeronautics and Space Administration. This research is partially supported by NASA ROSES ATP 14-ATP14-0093 grant.
RdP and O.D. acknowledge support by the Heising-Simons foundation.


\appendix

\section{Parameters and Priors}
\label{app:priors}

In this Appendix, we consider parameter priors and other constraints assumed in the MCMC likelihood analysis. For each model, there is a set of basic parameter priors, given in Tables \ref{tb:params case A} - \ref{tb:params case C}. On top of these priors, various additional constraints, other than those from the data discussed in the main text, are imposed.
There can be significant redundancy in these priors and constraints, i.e.~they are not all independent.
Let us first consider requirements that are imposed on all three models.
\begin{itemize}
\item
First of all, we always demand the spectator definition given in Eq.~(\ref{eq:spec def}) is satisfied. Secondly, we require all four slow-roll parameters at $t_*$ to be small,
\beq
\epsilon^\phi_*, \epsilon^\chi_*, |\eta^\phi_*|, |\eta^\chi_*| < 0.1.
\eeq
\item
Moreover, for a classical treatment of the spectator field to be appropriate, we
require that its initial value is much larger than the initial quantum fluctuations,
\beq
\chi_* > 10 \times \delta \chi_* = 10 \times \frac{H_*}{2 \pi}
\eeq
(we use a fixed value $H_* = 4 \cdot 10^{-5} \, \mpl$).
\item
Unless otherwise noted, we apply logarithmic priors to dimensionful parameters (and parameters that were dimensionful before dividing out powers of $m_\phi$).
\item
We will define the spectator-dominated regime by the somewhat arbitrary threshold,
\beq
R > 0.9.
\eeq
\end{itemize}

Let us now consider the specific parameters and priors/constraints for each model.

\subsection{Priors Case A: Quadratic-Axion}
\label{subsec:app case A}

\begin{itemize}
\item
The parameters sampled in the MCMC and their default prior ranges are given in Table \ref{tb:params case A}.
\item
As already incorporated there, we assume by default that the decay constant is sub-Planckian
\beq
f < M_{\rm Pl}
\eeq
(note that this is the Planck mass, not the reduced Planck mass), although we explicitly study how the results depend on the upper bound.
\item
We impose a linear prior on $\chi_*/f$ because in axion models, this quantity arises as a random phase.
\item
The lower bound on $V_0/m_\phi^2$ is derived from the requirement that reheating occurs at an energy
$\rho_{\rm reh} = 3 \mpl^2 \, H_{\rm reh}^2 > (10^3 \, $GeV$)^4$ (cf.~\cite{planck2015}),
i.e.~before the electroweak phase transition,
and that before that time the constant-$\zeta$ phase is reached where both fields are oscillating around their potential minima (see Fig.~\ref{fig:QAdens}).
Since the spectator/axion starts rolling approximately\footnote{We have confirmed the dependence of the time $\chi$ starts rolling on $V_0$ numerically.} when its mass $m_\chi$, given by Eq.~\eqref{mchiV0},
exceeds the Hubble scale $H$, we obtain an order-of-magnitude lower bound (assuming $m_\phi \sim 10^{13}$ GeV) of
$(V_0/m_{\phi}^2)_{\rm min}=10^{-52} \, \mpl^2$.

\end{itemize}

\renewcommand{\arraystretch}{1.4}
\begin{table}[t]
\small
\begin{center}
\begin{adjustbox}{max width=\textwidth}
\begin{tabular}{|l|l|c|}
\hline
Param. & Description &  Prior \\
\hline \hline
$V_0/m_\phi^2$ &  spectator amplitude & $[10^{-52} \mpl^2,10^5 \mpl^2]$ (log)  \\
$f$ &  spectator ``decay constant'' & $[10^{-4} \mpl, M_{\rm Pl}]$ (log)  \\
$\chi_*/f$ &  spectator initial phase & $[0,1/2]$ (linear)  \\
$\phi_*$ &  inflaton initial field & $[10 \mpl,30 \mpl]$ (log)  \\
\hline
\end{tabular}
\end{adjustbox}
\end{center}
\caption{Parameters and default priors for Case A: the quadratic-axion model. Additional constraints on the parameters are described in the text.}
\label{tb:params case A}
\end{table}

\subsection{Priors Case B: Curvaton}

\renewcommand{\arraystretch}{1.4}
\begin{table}[t]
\small
\begin{center}
\begin{adjustbox}{max width=\textwidth}
\begin{tabular}{|l|l|c|}
\hline
Param. & Description &  Prior \\
\hline \hline
$m_\chi/m_\phi$ &  spectator mass & $[10^{-26}, 1]$ (log)  \\
$\Gamma_{\rm reh}/m_\phi$ &  spectator reheating rate  & $[10^{-26}, 1]$ (log)  \\
$\chi_*$ &  spectator initial field & $[10 \times H_*/2\pi, \mpl]$ (log)  \\
$\phi_*$ &  inflaton initial field & $[3 \mpl,35 \mpl]$ (log)  \\
\hline
\end{tabular}
\end{adjustbox}
\end{center}
\caption{Parameters and default priors for Case B: the curvaton model. Additional constraints on the parameters are described in the text.}
\label{tb:params case B}
\end{table}

\begin{itemize}
\item
The parameters sampled in the MCMC and their default prior ranges are given in Table \ref{tb:params case B}.
\item
The curvaton scenario under consideration requires
\beq
m_\phi > m_\chi > \Gamma_{\rm reh}.
\eeq
\item
As discussed in the main text, we require the curvaton to be subdominant up to $t_{\rm curv}$,
\beq
\tfrac{1}{2} m_\chi^2 \, \chi_{\rm curv}^2 \ll 3 m_\chi^2.
\eeq
\item
Finally, we have determined the minimum value of $\Gamma_{\rm reh}$ in Table \ref{tb:params case A} as in case A, requiring $\rho_{\rm reh} > (10^3 \, $GeV$)^4$. This leads to the prior,
\beq
\frac{\Gamma_{\rm reh}}{m_\phi} > 10^{-26}.
\eeq

\end{itemize}

\subsection{Priors Case C: Modulated Reheating}

\renewcommand{\arraystretch}{1.4}
\begin{table}[t]
\small
\begin{center}
\begin{adjustbox}{max width=\textwidth}
\begin{tabular}{|l||l|c|}
\hline
Param. & Description &  Prior \\
\hline \hline
$m_\chi/m_\phi$ &  spectator mass & $[10^{-26}, 1]$ (log)  \\
$\lambda_0$ &  reheating coupling parameter & $[0, 1/2]$ (linear)  \\
$\lambda_1$ &  reheating coupling parameter & $[0, 1]$ (linear)  \\
$\lambda_2$ &  reheating coupling parameter & $[0, 1]$ (linear)  \\
$M_c$ & reheating cutoff parameter & $[10 \times H_*, \mpl]$ (log)  \\
$\chi_*$ &  spectator initial field & $[10 \times H_*/2\pi, \mpl]$ (log)  \\
$\phi_*$ &  inflaton initial field & $[3 \mpl,35 \mpl]$ (log)  \\
\hline
\end{tabular}
\end{adjustbox}
\end{center}
\caption{Parameters and default priors for Case C: the modulated reheating model. Additional constraints on the parameters are described in the text.}
\label{tb:params case C}
\end{table}

\begin{itemize}
\item
The parameters sampled in the MCMC and their default prior ranges are given in Table \ref{tb:params case C}.
\item
The specific scenario under consideration corresponds to the requirement,
\beq
m_\phi > \Gamma_{\rm reh}(\chi_{\rm reh}) > m_\chi.
\eeq
\item
We also impose
\beq
\frac{m_\chi}{m_\phi} > 10^{-26},
\eeq
which ensures reheating takes place before the electroweak phase transition (because $\Gamma_{\rm reh} > m_\chi$).
\item
We require the coupling constant, Eq.~(\ref{eq:coupling}), and its individual contributions to be significantly smaller than one\footnote{This does not ensure that non-perturbative effects are unimportant in the reheating process, but simply that the coupling constant is small enough for the leading-order perturbation theory expression for $\Gamma_{\rm reh}$ to be valid. In addition to this, there may well be non-perturbative effects even for small $\lambda$, such as resonant particle production from the vacuum.} (to rule out cases where the individual terms are large but cancel due to opposite signs),
\beq
|\lambda| < 1/2, \quad |\lambda_1| \, \frac{\chi_*}{M_c} < 1/2, \quad \tfrac{1}{2} |\lambda_2| \, \left( \frac{\chi_*}{M_c} \right)^2 < 1/2.
\eeq

\item
Finally, as already included in Table \ref{tb:params case C}, we demand that the cutoff $M_c$ is significantly above the Hubble scale at $t_*$,
\beq
M_c > 10 \times H_*.
\eeq
In practice, we incorporate $M_c$ into redefinitions of $\lambda_1$ and $\lambda_2$ and marginalize $M_c$ out analytically.
\end{itemize}

\section{Testing the Horizon Crossing Approximation}
\label{app:HCA}

We discuss here the accuracy of the Horizon Crossing Approximation for computing the curvature perturbations and $\fnl$ in the quadratic-axion model.
To do so, we numerically compute $\fnl$ in the $\delta N$ formalism using the full equations of motion, i.e.
\bea
\ddot \phi + 3 H \dot{\phi} + U_\phi  &=& 0 \nonumber \\
\ddot \chi + 3 H \dot{\chi} + V_\chi &=& 0 \nonumber \\
3 \mpl^2 \, H^2 &=& \tfrac{1}{2} \dot{\phi}^2 + \tfrac{1}{2} \dot{\chi}^2 + U(\phi) + V(\chi).
\eea

We compute the number of $e$-folds $N$ to a constant energy density hypersurface at a time when both fields have started oscillating around their minima, so that $\zeta$ has become constant. From there, one can get the numerical derivatives of $N$ and compute $\fnl$ using Eq.~\eqref{eq:fnlgeneral}.

As discussed in the main text, if we do not impose the spectator domination condition, the posterior is dominated by
points in parameter space where $\chi_*/f$ is not tuned to be small, so that the inflaton dominates the final perturbations and $\fnl \approx 0$.
Our main region of interest for testing the HCA is thus the regime where we explicitly impose the spectator domination condition,
\beq
\label{eq:spec-dom req}
\frac{\chi_*}{f} < \frac{f}{3 \pi^2 \phi_*}.
\eeq
This condition in turn favors larger values of $f$ as they leave a larger range of initial field values that satisfy the above requirement. We thus mainly want to test the HCA for models with $f$ within, say, an order of magnitude from the cutoff, i.e.~$f$ close to the Planck scale $M_{\rm pl}$.

\begin{figure}[h!]
\centering
\includegraphics[width=.49\textwidth]{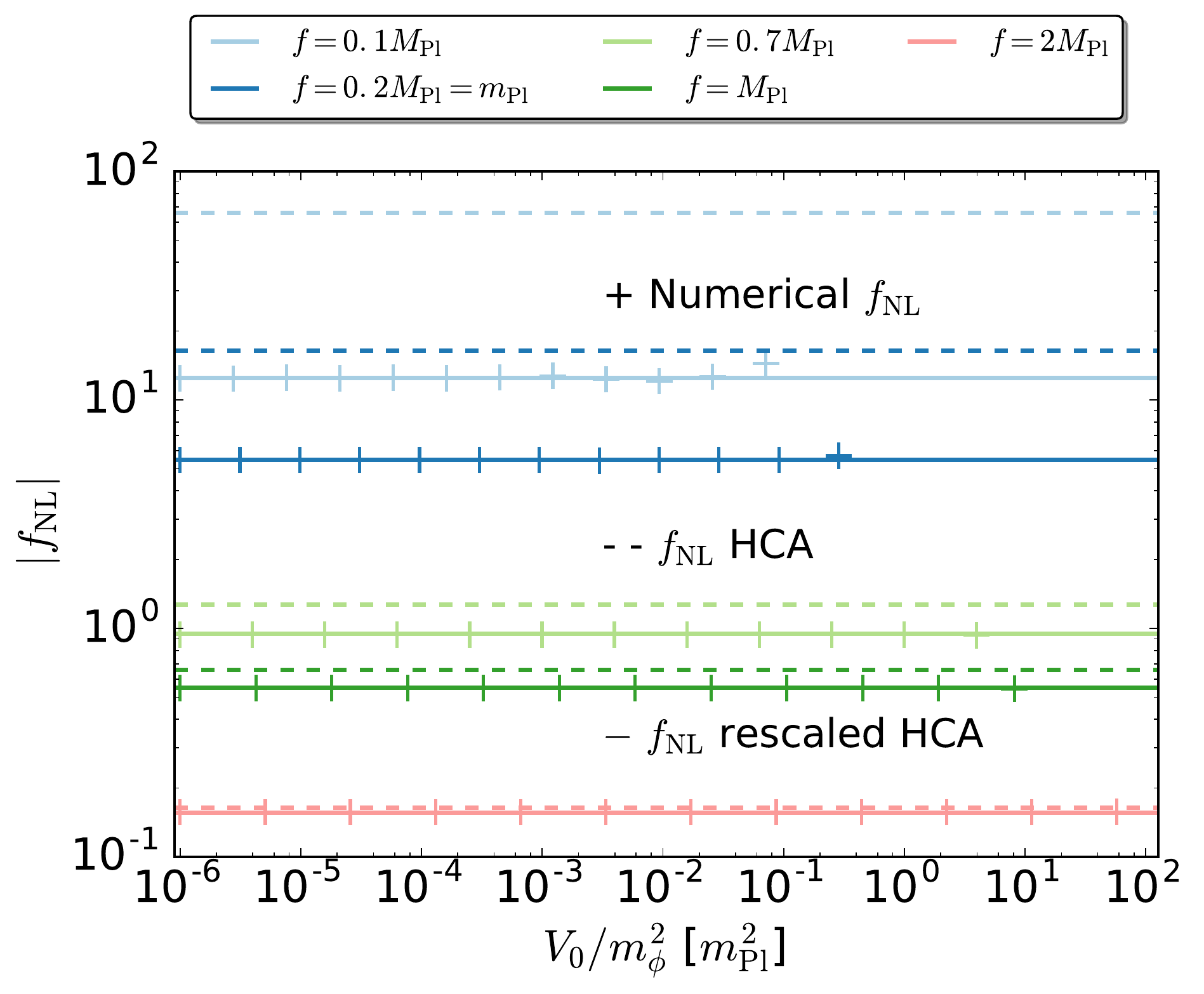}
\caption{Comparison of different approaches to calculating $\fnl$ in the quadratic-axion model (Case A).
We show $|\fnl|$ as a function of $V_0/m_{\phi}^2$ using the Horizon Crossing Approximation (dashed lines), the HCA with $f$-dependent correction factor (straight lines) and the full numerical calculation in the $\delta N$ formalism (+).
Results are computed in the limit $\chi_*/f \to 0$ (see text) and we restrict the numerical computation to the range of $V_0/m_\phi^2$ values relevant for our likelihood analysis ($|\eta^\chi_*| < 0.1$).
While accurate at large $f$, the HCA is only correct to order-of-magnitude level precision for $f$ significantly below $M_{\rm Pl}$. We have used the numerical calculations to construct an $f$-dependent rescaling function that brings the HCA prediction in good agreement with the exact result.}
\label{fig:numericsHCA}
\end{figure}

In Figure \ref{fig:numericsHCA}, we show $\fnl$ as a function of the amplitude of the axion potential, $V_0/m_\phi^2$, for various values of $f$ in the range motivated above.
For each parameter choice, we find that $\fnl$ converges as $\chi_*/f \to 0$ and in the plot we have chosen
values $\chi_*/f \ll f/(3 \pi^2 \phi_*)$ such that convergence has been reached.
The results are minimally sensitive to the choice of $\phi_*$.

For large values of $f$, the HCA (dashed lines) is a reasonably good approximation to the exact numerical results (plus signs).
For smaller $f$ however, the HCA systematically overpredicts $\fnl$.
For comparison, $|\fnl| < 5$ (cf.~Figure \ref{fig:fnlQuadAxion_simple}) corresponds to $f \gtrsim 0.4 M_{\rm pl}$, between the blue and light green results in Figure \ref{fig:numericsHCA}.
In order to account for the difference between the HCA and exact result, we implemented a simple function of $f$ that rescales $\fnl$ to make it agree with the numerical results. We show the new $\fnl$ with this correction factor in solid lines.
With the correction factor included, the agreement is quite good, except at high $V_0/m_\phi^2$ and low $f$ (again, the low $f$ regime is less relevant in our MCMC analysis).
We note that the results do depend also on $\chi_*/f$. While here we have shown the results in the low $\chi_*/f \to 0$ limit,
for values of $\chi_*/f$ that marginally satisfy Eq.~(\ref{eq:spec-dom req}),
we find a deviation from the results plotted here. However, the results are always within the range set by the HCA approximation (dashed) and the HCA approximation modified with the correction factor (solid).
In our likelihood analysis, to bracket the range of $\fnl$ values, we have considered results using either prescription.

We have also looked at how other quantities, such as $n_s$, $r$ and $R$ are different when using the full equations. Those differences are much smaller, and if the HCA values are within the Planck constraints, so are the ones from the full equations.
Moreover, since our main focus is primordial non-Gaussianity, the specifics of those other parameters are of lesser importance to our analysis. Indeed, as we mentioned before, they could be adjusted by a different choice of potential for the inflaton.

\newpage
\bibliographystyle{utphys}
\bibliography{refs}

\providecommand{\href}[2]{#2}\begingroup\raggedright\begin{thebibliography}{10}

\bibitem{staro80}
A.~A. {Starobinsky}, ``{A new type of isotropic cosmological models without
  singularity},'' \href{http://dx.doi.org/10.1016/0370-2693(80)90670-X}{{\em
  Physics Letters B} {\bfseries 91} (Mar., 1980) 99--102}.

\bibitem{guth81}
A.~H. {Guth}, ``{Inflationary universe: A possible solution to the horizon and
  flatness problems},'' \href{http://dx.doi.org/10.1103/PhysRevD.23.347}{{\em
  \prd} {\bfseries 23} (Jan., 1981) 347--356}.

\bibitem{linde82}
A.~D. {Linde}, ``{A new inflationary universe scenario: A possible solution of
  the horizon, flatness, homogeneity, isotropy and primordial monopole
  problems},'' \href{http://dx.doi.org/10.1016/0370-2693(82)91219-9}{{\em
  Physics Letters B} {\bfseries 108} (Feb., 1982) 389--393}.

\bibitem{albrechtstein82}
A.~{Albrecht} and P.~J. {Steinhardt}, ``{Cosmology for grand unified theories
  with radiatively induced symmetry breaking},''
  \href{http://dx.doi.org/10.1103/PhysRevLett.48.1220}{{\em Physical Review
  Letters} {\bfseries 48} (Apr., 1982) 1220--1223}.

\bibitem{cvrcekwitten06}
P.~{Svrcek} and E.~{Witten}, ``{Axions in string theory},''
  \href{http://dx.doi.org/10.1088/1126-6708/2006/06/051}{{\em Journal of High
  Energy Physics} {\bfseries 6} (June, 2006) 051},
  \href{http://arxiv.org/abs/hep-th/0605206}{{\ttfamily hep-th/0605206}}.

\bibitem{dimoetal08}
S.~{Dimopoulos}, S.~{Kachru}, J.~{McGreevy}, {\em et~al.}, ``{N-flation},''
  \href{http://dx.doi.org/10.1088/1475-7516/2008/08/003}{{\em \jcap} {\bfseries
  8} (Aug., 2008) 003}, \href{http://arxiv.org/abs/hep-th/0507205}{{\ttfamily
  hep-th/0507205}}.

\bibitem{baumc15}
D.~{Baumann} and L.~{McAllister}, {\em {Inflation and String Theory}}.
\newblock Apr., 2015.

\bibitem{komatsuetal05}
E.~{Komatsu}, D.~N. {Spergel}, and B.~D. {Wandelt}, ``{Measuring Primordial
  Non-Gaussianity in the Cosmic Microwave Background},''
  \href{http://dx.doi.org/10.1086/491724}{{\em \apj} {\bfseries 634} (Nov.,
  2005) 14--19}, \href{http://arxiv.org/abs/astro-ph/0305189}{{\ttfamily
  astro-ph/0305189}}.

\bibitem{Maldacena:2002vr}
J.~M. Maldacena, ``{Non-Gaussian Features of Primordial Fluctuations in
  Single-Field Inflationary Models},''
  \href{http://dx.doi.org/10.1088/1126-6708/2003/05/013}{{\em JHEP} {\bfseries
  0305} (2003) 013},
\href{http://arxiv.org/abs/astro-ph/0210603}{{\ttfamily arXiv:astro-ph/0210603
  [astro-ph]}}.

\bibitem{Creminelli:2004yq}
P.~Creminelli and M.~Zaldarriaga, ``{Single field consistency relation for the
  3-point function},''
  \href{http://dx.doi.org/10.1088/1475-7516/2004/10/006}{{\em JCAP} {\bfseries
  0410} (2004) 006},
\href{http://arxiv.org/abs/astro-ph/0407059}{{\ttfamily arXiv:astro-ph/0407059
  [astro-ph]}}.

\bibitem{chenetal13}
X.~{Chen}, H.~{Firouzjahi}, M.~H. {Namjoo}, {\em et~al.}, ``{A single field
  inflation model with large local non-Gaussianity},''
  \href{http://dx.doi.org/10.1209/0295-5075/102/59001}{{\em EPL (Europhysics
  Letters)} {\bfseries 102} (June, 2013) 59001},
  \href{http://arxiv.org/abs/1301.5699}{{\ttfamily arXiv:1301.5699 [hep-th]}}.

\bibitem{mooijpalma15}
S.~{Mooij} and G.~A. {Palma}, ``{Consistently violating the non-Gaussian
  consistency relation},''
  \href{http://dx.doi.org/10.1088/1475-7516/2015/11/025}{{\em \jcap} {\bfseries
  11} (Nov., 2015) 025}, \href{http://arxiv.org/abs/1502.03458}{{\ttfamily
  arXiv:1502.03458}}.

\bibitem{byrneschoi10}
C.~T. {Byrnes} and K.-Y. {Choi}, ``{Review of Local Non-Gaussianity from
  Multifield Inflation},'' \href{http://dx.doi.org/10.1155/2010/724525}{{\em
  Advances in Astronomy} {\bfseries 2010} (2010) 724525},
  \href{http://arxiv.org/abs/1002.3110}{{\ttfamily arXiv:1002.3110}}.

\bibitem{planck2015png}
{Planck Collaboration}, P.~A.~R. {Ade}, N.~{Aghanim}, {\em et~al.}, ``{Planck
  2015 results. XVII. Constraints on primordial non-Gaussianity},'' {\em ArXiv
  e-prints} (Feb., 2015) , \href{http://arxiv.org/abs/1502.01592}{{\ttfamily
  arXiv:1502.01592}}.

\bibitem{toronto14}
M.~{Alvarez}, T.~{Baldauf}, J.~R. {Bond}, {\em et~al.}, ``{Testing Inflation
  with Large Scale Structure: Connecting Hopes with Reality},'' {\em ArXiv
  e-prints} (Dec., 2014) , \href{http://arxiv.org/abs/1412.4671}{{\ttfamily
  arXiv:1412.4671}}.

\bibitem{Dalal:2007cu}
N.~Dalal, O.~Dor{\'e}, D.~Huterer, {\em et~al.}, ``{The imprints of primordial
  non-gaussianities on large-scale structure: scale dependent bias and
  abundance of virialized objects},''
  \href{http://dx.doi.org/10.1103/PhysRevD.77.123514}{{\em Phys.Rev.}
  {\bfseries D77} (2008) 123514},
\href{http://arxiv.org/abs/0710.4560}{{\ttfamily arXiv:0710.4560 [astro-ph]}}.

\bibitem{matver08}
S.~{Matarrese} and L.~{Verde}, ``{The Effect of Primordial Non-Gaussianity on
  Halo Bias},'' \href{http://dx.doi.org/10.1086/587840}{{\em \apjl} {\bfseries
  677} (Apr., 2008) L77}, \href{http://arxiv.org/abs/0801.4826}{{\ttfamily
  arXiv:0801.4826}}.

\bibitem{slosaretal08}
A.~{Slosar}, C.~{Hirata}, U.~{Seljak}, {\em et~al.}, ``{Constraints on local
  primordial non-Gaussianity from large scale structure},''
  \href{http://dx.doi.org/10.1088/1475-7516/2008/08/031}{{\em \jcap} {\bfseries
  8} (Aug., 2008) 031}, \href{http://arxiv.org/abs/0805.3580}{{\ttfamily
  arXiv:0805.3580}}.

\bibitem{desjseljak10}
V.~{Desjacques} and U.~{Seljak}, ``{Primordial non-Gaussianity from the
  large-scale structure},''
  \href{http://dx.doi.org/10.1088/0264-9381/27/12/124011}{{\em Classical and
  Quantum Gravity} {\bfseries 27} no.~12, (June, 2010) 124011},
  \href{http://arxiv.org/abs/1003.5020}{{\ttfamily arXiv:1003.5020}}.

\bibitem{spherex_wp}
O.~{Dor{\'e}}, J.~{Bock}, M.~{Ashby}, {\em et~al.}, ``{Cosmology with the
  SPHEREX All-Sky Spectral Survey},'' {\em ArXiv e-prints} (Dec., 2014) ,
  \href{http://arxiv.org/abs/1412.4872}{{\ttfamily arXiv:1412.4872}}.

\bibitem{Dore:2016tfs}
O.~Doré {\em et~al.}, ``{Science Impacts of the SPHEREx All-Sky Optical to
  Near-Infrared Spectral Survey: Report of a Community Workshop Examining
  Extragalactic, Galactic, Stellar and Planetary Science},''
\href{http://arxiv.org/abs/1606.07039}{{\ttfamily arXiv:1606.07039
  [astro-ph.CO]}}.

\bibitem{Spherexweb}
\url{http://spherex.caltech.edu/}.

\bibitem{lsst}
{LSST Science Collaboration}, P.~A. {Abell}, J.~{Allison}, {\em et~al.},
  ``{LSST Science Book, Version 2.0},'' {\em ArXiv e-prints} (Dec., 2009) ,
  \href{http://arxiv.org/abs/0912.0201}{{\ttfamily arXiv:0912.0201
  [astro-ph.IM]}}.

\bibitem{euclid}
R.~{Laureijs}, J.~{Amiaux}, S.~{Arduini}, {\em et~al.}, ``{Euclid Definition
  Study Report},'' {\em ArXiv e-prints} (Oct., 2011) ,
  \href{http://arxiv.org/abs/1110.3193}{{\ttfamily arXiv:1110.3193
  [astro-ph.CO]}}.

\bibitem{S4}
K.~N. {Abazajian}, P.~{Adshead}, Z.~{Ahmed}, {\em et~al.}, ``{CMB-S4 Science
  Book, First Edition},'' {\em ArXiv e-prints} (Oct., 2016) ,
  \href{http://arxiv.org/abs/1610.02743}{{\ttfamily arXiv:1610.02743}}.

\bibitem{RdPOli14}
R.~{de Putter} and O.~{Dor{\'e}}, ``{Designing an Inflation Galaxy Survey: how
  to measure $\sigma(\fnl) \sim 1$ using scale-dependent galaxy bias},'' {\em
  ArXiv e-prints} (Dec., 2014) ,
  \href{http://arxiv.org/abs/1412.3854}{{\ttfamily arXiv:1412.3854}}.

\bibitem{ferrsmith14}
S.~{Ferraro} and K.~M. {Smith}, ``{Using Large Scale Structure to test
  Multifield Inflation},'' {\em ArXiv e-prints} (Aug., 2014) ,
  \href{http://arxiv.org/abs/1408.3126}{{\ttfamily arXiv:1408.3126}}.

\bibitem{yamauchietal14}
D.~{Yamauchi}, K.~{Takahashi}, and M.~{Oguri}, ``{Constraining primordial
  non-Gaussianity via a multitracer technique with surveys by Euclid and the
  Square Kilometre Array},''
  \href{http://dx.doi.org/10.1103/PhysRevD.90.083520}{{\em \prd} {\bfseries 90}
  no.~8, (Oct., 2014) 083520}, \href{http://arxiv.org/abs/1407.5453}{{\ttfamily
  arXiv:1407.5453}}.

\bibitem{Byrnes:2008wi}
C.~T. Byrnes, K.-Y. Choi, and L.~M.~H. Hall, ``{Conditions for large
  non-Gaussianity in two-field slow-roll inflation},''
  \href{http://dx.doi.org/10.1088/1475-7516/2008/10/008}{{\em JCAP} {\bfseries
  0810} (2008) 008},
\href{http://arxiv.org/abs/0807.1101}{{\ttfamily arXiv:0807.1101 [astro-ph]}}.

\bibitem{lindemukh97}
A.~{Linde} and V.~{Mukhanov}, ``{Non-Gaussian isocurvature perturbations from
  inflation},'' \href{http://dx.doi.org/10.1103/PhysRevD.56.R535}{{\em \prd}
  {\bfseries 56} (July, 1997) R535--539},
  \href{http://arxiv.org/abs/astro-ph/9610219}{{\ttfamily astro-ph/9610219}}.

\bibitem{Moroi:2002rd}
T.~Moroi and T.~Takahashi, ``{Cosmic density perturbations from late decaying
  scalar condensations},''
  \href{http://dx.doi.org/10.1103/PhysRevD.66.063501}{{\em Phys. Rev.}
  {\bfseries D66} (2002) 063501},
\href{http://arxiv.org/abs/hep-ph/0206026}{{\ttfamily arXiv:hep-ph/0206026
  [hep-ph]}}.

\bibitem{lythetal03}
D.~H. {Lyth}, C.~{Ungarelli}, and D.~{Wands}, ``{Primordial density
  perturbation in the curvaton scenario},''
  \href{http://dx.doi.org/10.1103/PhysRevD.67.023503}{{\em \prd} {\bfseries 67}
  no.~2, (Jan., 2003) 023503},
  \href{http://arxiv.org/abs/astro-ph/0208055}{{\ttfamily astro-ph/0208055}}.

\bibitem{Lyth:2001nq}
D.~H. Lyth and D.~Wands, ``{Generating the curvature perturbation without an
  inflaton},'' \href{http://dx.doi.org/10.1016/S0370-2693(01)01366-1}{{\em
  Phys. Lett.} {\bfseries B524} (2002) 5--14},
\href{http://arxiv.org/abs/hep-ph/0110002}{{\ttfamily arXiv:hep-ph/0110002
  [hep-ph]}}.

\bibitem{Enqvist:2001zp}
K.~Enqvist and M.~S. Sloth, ``{Adiabatic CMB perturbations in pre - big bang
  string cosmology},''
  \href{http://dx.doi.org/10.1016/S0550-3213(02)00043-3}{{\em Nucl. Phys.}
  {\bfseries B626} (2002) 395--409},
\href{http://arxiv.org/abs/hep-ph/0109214}{{\ttfamily arXiv:hep-ph/0109214
  [hep-ph]}}.

\bibitem{Dvali:2003em}
G.~Dvali, A.~Gruzinov, and M.~Zaldarriaga, ``{A new mechanism for generating
  density perturbations from inflation},''
  \href{http://dx.doi.org/10.1103/PhysRevD.69.023505}{{\em Phys. Rev.}
  {\bfseries D69} (2004) 023505},
\href{http://arxiv.org/abs/astro-ph/0303591}{{\ttfamily arXiv:astro-ph/0303591
  [astro-ph]}}.

\bibitem{Kofman:2003nx}
L.~Kofman, ``{Probing string theory with modulated cosmological
  fluctuations},''
\href{http://arxiv.org/abs/astro-ph/0303614}{{\ttfamily arXiv:astro-ph/0303614
  [astro-ph]}}.

\bibitem{Bernardeau:2004zz}
F.~Bernardeau, L.~Kofman, and J.-P. Uzan, ``{Modulated fluctuations from hybrid
  inflation},'' \href{http://dx.doi.org/10.1103/PhysRevD.70.083004}{{\em Phys.
  Rev.} {\bfseries D70} (2004) 083004},
\href{http://arxiv.org/abs/astro-ph/0403315}{{\ttfamily arXiv:astro-ph/0403315
  [astro-ph]}}.

\bibitem{dvalietal04}
G.~{Dvali}, A.~{Gruzinov}, and M.~{Zaldarriaga}, ``{New mechanism for
  generating density perturbations from inflation},''
  \href{http://dx.doi.org/10.1103/PhysRevD.69.023505}{{\em \prd} {\bfseries 69}
  no.~2, (Jan., 2004) 023505},
  \href{http://arxiv.org/abs/astro-ph/0303591}{{\ttfamily astro-ph/0303591}}.

\bibitem{zal04}
M.~{Zaldarriaga}, ``{Non-Gaussianities in models with a varying inflaton decay
  rate},'' \href{http://dx.doi.org/10.1103/PhysRevD.69.043508}{{\em \prd}
  {\bfseries 69} no.~4, (Feb., 2004) 043508},
  \href{http://arxiv.org/abs/astro-ph/0306006}{{\ttfamily astro-ph/0306006}}.

\bibitem{bernardeauetal04}
F.~{Bernardeau}, L.~{Kofman}, and J.-P. {Uzan}, ``{Modulated fluctuations from
  hybrid inflation},'' \href{http://dx.doi.org/10.1103/PhysRevD.70.083004}{{\em
  \prd} {\bfseries 70} no.~8, (Oct., 2004) 083004},
  \href{http://arxiv.org/abs/astro-ph/0403315}{{\ttfamily astro-ph/0403315}}.

\bibitem{lyth05}
D.~H. {Lyth}, ``{Generating the curvature perturbation at the end of
  inflation},'' \href{http://dx.doi.org/10.1088/1475-7516/2005/11/006}{{\em
  \jcap} {\bfseries 11} (Dec., 2005) 006},
  \href{http://arxiv.org/abs/astro-ph/0510443}{{\ttfamily astro-ph/0510443}}.

\bibitem{huang09}
Q.-G. {Huang}, ``{A geometric description of the non-Gaussianity generated at
  the end of multi-field inflation},''
  \href{http://dx.doi.org/10.1088/1475-7516/2009/06/035}{{\em \jcap} {\bfseries
  6} (June, 2009) 035}, \href{http://arxiv.org/abs/0904.2649}{{\ttfamily
  arXiv:0904.2649 [hep-th]}}.

\bibitem{kawasakietal09}
M.~{Kawasaki}, T.~{Takahashi}, and S.~{Yokoyama}, ``{Density fluctuations in
  thermal inflation and non-Gaussianity},''
  \href{http://dx.doi.org/10.1088/1475-7516/2009/12/012}{{\em \jcap} {\bfseries
  12} (Dec., 2009) 012}, \href{http://arxiv.org/abs/0910.3053}{{\ttfamily
  arXiv:0910.3053 [hep-th]}}.

\bibitem{Elliston:2011dr}
J.~Elliston, D.~J. Mulryne, D.~Seery, {\em et~al.}, ``{Evolution of fNL to the
  adiabatic limit},''
  \href{http://dx.doi.org/10.1088/1475-7516/2011/11/005}{{\em JCAP} {\bfseries
  1111} (2011) 005},
\href{http://arxiv.org/abs/1106.2153}{{\ttfamily arXiv:1106.2153
  [astro-ph.CO]}}.

\bibitem{leungetal12}
G.~{Leung}, E.~R.~M. {Tarrant}, C.~T. {Byrnes}, {\em et~al.}, ``{Reheating,
  multifield inflation and the fate of the primordial observables},''
  \href{http://dx.doi.org/10.1088/1475-7516/2012/09/008}{{\em \jcap} {\bfseries
  9} (Sept., 2012) 008}, \href{http://arxiv.org/abs/1206.5196}{{\ttfamily
  arXiv:1206.5196 [astro-ph.CO]}}.

\bibitem{ellistonetal14}
J.~{Elliston}, S.~{Orani}, and D.~J. {Mulryne}, ``{General analytic predictions
  of two-field inflation and perturbative reheating},''
  \href{http://dx.doi.org/10.1103/PhysRevD.89.103532}{{\em \prd} {\bfseries 89}
  no.~10, (May, 2014) 103532}, \href{http://arxiv.org/abs/1402.4800}{{\ttfamily
  arXiv:1402.4800}}.

\bibitem{byrnesetal09}
C.~T. {Byrnes}, K.-Y. {Choi}, and L.~M.~H. {Hall}, ``{Large non-Gaussianity
  from two-component hybrid inflation},''
  \href{http://dx.doi.org/10.1088/1475-7516/2009/02/017}{{\em \jcap} {\bfseries
  2} (Feb., 2009) 017}, \href{http://arxiv.org/abs/0812.0807}{{\ttfamily
  arXiv:0812.0807}}.

\bibitem{narukosasaki09}
A.~{Naruko} and M.~{Sasaki}, ``{Large Non-Gaussianity from Multi-Brid
  Inflation},'' \href{http://dx.doi.org/10.1143/PTP.121.193}{{\em Progress of
  Theoretical Physics} {\bfseries 121} (Jan., 2009) 193--210},
  \href{http://arxiv.org/abs/0807.0180}{{\ttfamily arXiv:0807.0180}}.

\bibitem{Kim:2006te}
S.~A. Kim and A.~R. Liddle, ``{Nflation: Non-Gaussianity in the
  horizon-crossing approximation},''
  \href{http://dx.doi.org/10.1103/PhysRevD.74.063522}{{\em Phys. Rev.}
  {\bfseries D74} (2006) 063522},
\href{http://arxiv.org/abs/astro-ph/0608186}{{\ttfamily arXiv:astro-ph/0608186
  [astro-ph]}}.

\bibitem{kimetal10}
S.~A. {Kim}, A.~R. {Liddle}, and D.~{Seery}, ``{Non-Gaussianity in Axion
  N-flation Models},''
  \href{http://dx.doi.org/10.1103/PhysRevLett.105.181302}{{\em Physical Review
  Letters} {\bfseries 105} no.~18, (Oct., 2010) 181302},
  \href{http://arxiv.org/abs/1005.4410}{{\ttfamily arXiv:1005.4410}}.

\bibitem{langloissorbo09}
D.~{Langlois} and L.~{Sorbo}, ``{Primordial perturbations and non-Gaussianities
  from modulated trapping},''
  \href{http://dx.doi.org/10.1088/1475-7516/2009/08/014}{{\em \jcap} {\bfseries
  8} (Aug., 2009) 014}, \href{http://arxiv.org/abs/0906.1813}{{\ttfamily
  arXiv:0906.1813 [astro-ph.CO]}}.

\bibitem{nakayamasuyama11}
K.~{Nakayama} and T.~{Suyama}, ``{Curvature perturbation from velocity
  modulation},'' \href{http://dx.doi.org/10.1103/PhysRevD.84.063520}{{\em \prd}
  {\bfseries 84} no.~6, (Sept., 2011) 063520},
  \href{http://arxiv.org/abs/1107.3003}{{\ttfamily arXiv:1107.3003
  [astro-ph.CO]}}.

\bibitem{petersontegmark11a}
C.~M. {Peterson} and M.~{Tegmark}, ``{Non-Gaussianity in two-field
  inflation},'' \href{http://dx.doi.org/10.1103/PhysRevD.84.023520}{{\em \prd}
  {\bfseries 84} no.~2, (July, 2011) 023520},
  \href{http://arxiv.org/abs/1011.6675}{{\ttfamily arXiv:1011.6675}}.

\bibitem{petersontegmark11b}
C.~M. {Peterson} and M.~{Tegmark}, ``{Testing Multi-Field Inflation: A
  Geometric Approach},'' {\em ArXiv e-prints} (Nov., 2011) ,
  \href{http://arxiv.org/abs/1111.0927}{{\ttfamily arXiv:1111.0927
  [astro-ph.CO]}}.

\bibitem{alabidietal10}
L.~{Alabidi}, K.~{Malik}, C.~T. {Byrnes}, {\em et~al.}, ``{How the curvaton
  scenario, modulated reheating and an inhomogeneous end of inflation are
  related},'' \href{http://dx.doi.org/10.1088/1475-7516/2010/11/037}{{\em
  \jcap} {\bfseries 11} (Nov., 2010) 037},
  \href{http://arxiv.org/abs/1002.1700}{{\ttfamily arXiv:1002.1700
  [astro-ph.CO]}}.

\bibitem{byrnesetal14}
C.~T. {Byrnes}, M.~{Cort{\^e}s}, and A.~R. {Liddle}, ``{Comprehensive analysis
  of the simplest curvaton model},''
  \href{http://dx.doi.org/10.1103/PhysRevD.90.023523}{{\em \prd} {\bfseries 90}
  no.~2, (July, 2014) 023523}, \href{http://arxiv.org/abs/1403.4591}{{\ttfamily
  arXiv:1403.4591}}.

\bibitem{smithgrin15}
T.~L. {Smith} and D.~{Grin}, ``{Probing a panoply of curvaton-decay scenarios
  using CMB data},'' {\em ArXiv e-prints} (Nov., 2015) ,
  \href{http://arxiv.org/abs/1511.07431}{{\ttfamily arXiv:1511.07431}}.

\bibitem{venninetal15}
V.~{Vennin}, K.~{Koyama}, and D.~{Wands}, ``{Encyclop{\ae}dia curvatonis},''
  \href{http://dx.doi.org/10.1088/1475-7516/2015/11/008}{{\em \jcap} {\bfseries
  11} (Nov., 2015) 008}, \href{http://arxiv.org/abs/1507.07575}{{\ttfamily
  arXiv:1507.07575}}.

\bibitem{venninetal16}
V.~{Vennin}, K.~{Koyama}, and D.~{Wands}, ``{Inflation with an extra light
  scalar field after Planck},''
  \href{http://dx.doi.org/10.1088/1475-7516/2016/03/024}{{\em \jcap} {\bfseries
  3} (Mar., 2016) 024}, \href{http://arxiv.org/abs/1512.03403}{{\ttfamily
  arXiv:1512.03403}}.

\bibitem{staro85}
A.~A. {Starobinski{\v i}}, ``{Multicomponent de Sitter (inflationary) stages
  and the generation of perturbations},'' {\em Soviet Journal of Experimental
  and Theoretical Physics Letters} {\bfseries 42} (Aug., 1985) 152.

\bibitem{Sasaki:1995aw}
M.~Sasaki and E.~D. Stewart, ``{A General analytic formula for the spectral
  index of the density perturbations produced during inflation},''
  \href{http://dx.doi.org/10.1143/PTP.95.71}{{\em Prog. Theor. Phys.}
  {\bfseries 95} (1996) 71--78},
\href{http://arxiv.org/abs/astro-ph/9507001}{{\ttfamily arXiv:astro-ph/9507001
  [astro-ph]}}.

\bibitem{lythrodr05}
D.~H. {Lyth} and Y.~{Rodr{\'{\i}}guez}, ``{Inflationary Prediction for
  Primordial Non-Gaussianity},''
  \href{http://dx.doi.org/10.1103/PhysRevLett.95.121302}{{\em Physical Review
  Letters} {\bfseries 95} no.~12, (Sept., 2005) 121302},
  \href{http://arxiv.org/abs/astro-ph/0504045}{{\ttfamily astro-ph/0504045}}.

\bibitem{vernwands06}
F.~{Vernizzi} and D.~{Wands}, ``{Non-Gaussianities in two-field inflation},''
  \href{http://dx.doi.org/10.1088/1475-7516/2006/05/019}{{\em JCAP} {\bfseries
  5} (May, 2006) 19}, \href{http://arxiv.org/abs/astro-ph/0603799}{{\ttfamily
  astro-ph/0603799}}.

\bibitem{kobaetal13}
T.~{Kobayashi}, F.~{Takahashi}, T.~{Takahashi}, {\em et~al.}, ``{Spectator
  field models in light of spectral index after Planck},''
  \href{http://dx.doi.org/10.1088/1475-7516/2013/10/042}{{\em \jcap} {\bfseries
  10} (Oct., 2013) 042}, \href{http://arxiv.org/abs/1303.6255}{{\ttfamily
  arXiv:1303.6255 [astro-ph.CO]}}.

\bibitem{enqtak13}
K.~{Enqvist} and T.~{Takahashi}, ``{Mixed inflaton and spectator field models
  after Planck},'' \href{http://dx.doi.org/10.1088/1475-7516/2013/10/034}{{\em
  \jcap} {\bfseries 10} (Oct., 2013) 034},
  \href{http://arxiv.org/abs/1306.5958}{{\ttfamily arXiv:1306.5958}}.

\bibitem{wang16}
L.~{Wang}, ``{Spectator fields and their imprints on the Cosmic Microwave
  Background},'' {\em ArXiv e-prints} (Oct., 2016) ,
  \href{http://arxiv.org/abs/1610.03123}{{\ttfamily arXiv:1610.03123}}.

\bibitem{suyamaetal13}
T.~{Suyama}, T.~{Takahashi}, M.~{Yamaguchi}, {\em et~al.}, ``{Implications of
  Planck results for models with local type non-Gaussianity},''
  \href{http://dx.doi.org/10.1088/1475-7516/2013/06/012}{{\em \jcap} {\bfseries
  6} (June, 2013) 012}, \href{http://arxiv.org/abs/1303.5374}{{\ttfamily
  arXiv:1303.5374}}.

\bibitem{renpet15}
S.~{Renaux-Petel}, ``{Primordial non-Gaussianities after Planck 2015: An
  introductory review},''
  \href{http://dx.doi.org/10.1016/j.crhy.2015.08.003}{{\em Comptes Rendus
  Physique} {\bfseries 16} (Dec., 2015) 969--985},
  \href{http://arxiv.org/abs/1508.06740}{{\ttfamily arXiv:1508.06740}}.

\bibitem{Bardeen:1983qw}
J.~M. Bardeen, P.~J. Steinhardt, and M.~S. Turner, ``{Spontaneous Creation of
  Almost Scale - Free Density Perturbations in an Inflationary Universe},''
\href{http://dx.doi.org/10.1103/PhysRevD.28.679}{{\em Phys. Rev.} {\bfseries
  D28} (1983) 679}.

\bibitem{Wands:2000dp}
D.~Wands, K.~A. Malik, D.~H. Lyth, {\em et~al.}, ``{A New approach to the
  evolution of cosmological perturbations on large scales},''
  \href{http://dx.doi.org/10.1103/PhysRevD.62.043527}{{\em Phys. Rev.}
  {\bfseries D62} (2000) 043527},
\href{http://arxiv.org/abs/astro-ph/0003278}{{\ttfamily arXiv:astro-ph/0003278
  [astro-ph]}}.

\bibitem{Weinberg:2003sw}
S.~Weinberg, ``{Adiabatic modes in cosmology},''
  \href{http://dx.doi.org/10.1103/PhysRevD.67.123504}{{\em Phys.Rev.}
  {\bfseries D67} (2003) 123504},
\href{http://arxiv.org/abs/astro-ph/0302326}{{\ttfamily arXiv:astro-ph/0302326
  [astro-ph]}}.

\bibitem{Weinberg:2004kf}
S.~Weinberg, ``{Must cosmological perturbations remain non-adiabatic after
  multi-field inflation?},''
  \href{http://dx.doi.org/10.1103/PhysRevD.70.083522}{{\em Phys. Rev.}
  {\bfseries D70} (2004) 083522},
\href{http://arxiv.org/abs/astro-ph/0405397}{{\ttfamily arXiv:astro-ph/0405397
  [astro-ph]}}.

\bibitem{Weinberg:2008si}
S.~Weinberg, ``{Non-Gaussian Correlations Outside the Horizon II: The General
  Case},'' \href{http://dx.doi.org/10.1103/PhysRevD.79.043504}{{\em Phys. Rev.}
  {\bfseries D79} (2009) 043504},
\href{http://arxiv.org/abs/0810.2831}{{\ttfamily arXiv:0810.2831 [hep-ph]}}.

\bibitem{Meyers:2012ni}
J.~Meyers, ``{Non-Gaussian Correlations Outside the Horizon in Local Thermal
  Equilibrium},''
\href{http://arxiv.org/abs/1212.4438}{{\ttfamily arXiv:1212.4438
  [astro-ph.CO]}}.

\bibitem{Wands:2010af}
D.~Wands, ``{Local non-Gaussianity from inflation},''
  \href{http://dx.doi.org/10.1088/0264-9381/27/12/124002}{{\em Class. Quant.
  Grav.} {\bfseries 27} (2010) 124002},
\href{http://arxiv.org/abs/1004.0818}{{\ttfamily arXiv:1004.0818
  [astro-ph.CO]}}.

\bibitem{Sasaki:2006kq}
M.~Sasaki, J.~Valiviita, and D.~Wands, ``{Non-Gaussianity of the primordial
  perturbation in the curvaton model},''
  \href{http://dx.doi.org/10.1103/PhysRevD.74.103003}{{\em Phys. Rev.}
  {\bfseries D74} (2006) 103003},
\href{http://arxiv.org/abs/astro-ph/0607627}{{\ttfamily arXiv:astro-ph/0607627
  [astro-ph]}}.

\bibitem{Leung:2012ve}
G.~Leung, E.~R.~M. Tarrant, C.~T. Byrnes, {\em et~al.}, ``{Reheating,
  Multifield Inflation and the Fate of the Primordial Observables},''
  \href{http://dx.doi.org/10.1088/1475-7516/2012/09/008}{{\em JCAP} {\bfseries
  1209} (2012) 008},
\href{http://arxiv.org/abs/1206.5196}{{\ttfamily arXiv:1206.5196
  [astro-ph.CO]}}.

\bibitem{Meyers:2013gua}
J.~Meyers and E.~R.~M. Tarrant, ``{Perturbative Reheating After Multiple-Field
  Inflation: The Impact on Primordial Observables},''
  \href{http://dx.doi.org/10.1103/PhysRevD.89.063535}{{\em Phys. Rev.}
  {\bfseries D89} no.~6, (2014) 063535},
\href{http://arxiv.org/abs/1311.3972}{{\ttfamily arXiv:1311.3972
  [astro-ph.CO]}}.

\bibitem{Amin:2014eta}
M.~A. Amin, M.~P. Hertzberg, D.~I. Kaiser, {\em et~al.}, ``{Nonperturbative
  Dynamics Of Reheating After Inflation: A Review},''
  \href{http://dx.doi.org/10.1142/S0218271815300037}{{\em Int. J. Mod. Phys.}
  {\bfseries D24} (2014) 1530003},
\href{http://arxiv.org/abs/1410.3808}{{\ttfamily arXiv:1410.3808 [hep-ph]}}.

\bibitem{lidseyseery05}
D.~{Seery} and J.~E. {Lidsey}, ``{Primordial non-Gaussianities from
  multiple-field inflation},''
  \href{http://dx.doi.org/10.1088/1475-7516/2005/09/011}{{\em \jcap} {\bfseries
  9} (Sept., 2005) 11}, \href{http://arxiv.org/abs/astro-ph/0506056}{{\ttfamily
  astro-ph/0506056}}.

\bibitem{wandsetal02}
D.~{Wands}, N.~{Bartolo}, S.~{Matarrese}, {\em et~al.}, ``{Observational test
  of two-field inflation},''
  \href{http://dx.doi.org/10.1103/PhysRevD.66.043520}{{\em \prd} {\bfseries 66}
  no.~4, (Aug., 2002) 043520},
  \href{http://arxiv.org/abs/astro-ph/0205253}{{\ttfamily astro-ph/0205253}}.

\bibitem{alabidilyth06}
L.~{Alabidi} and D.~H. {Lyth}, ``{Inflation models and observation},''
  \href{http://dx.doi.org/10.1088/1475-7516/2006/05/016}{{\em \jcap} {\bfseries
  5} (May, 2006) 016}, \href{http://arxiv.org/abs/astro-ph/0510441}{{\ttfamily
  astro-ph/0510441}}.

\bibitem{kobayashietal14}
N.~{Kobayashi}, T.~{Kobayashi}, and A.~L. {Erickcek}, ``{Rolling in the
  modulated reheating scenario},''
  \href{http://dx.doi.org/10.1088/1475-7516/2014/01/036}{{\em \jcap} {\bfseries
  1} (Jan., 2014) 036}, \href{http://arxiv.org/abs/1308.4154}{{\ttfamily
  arXiv:1308.4154 [astro-ph.CO]}}.

\bibitem{meyerstarrant14}
J.~{Meyers} and E.~R.~M. {Tarrant}, ``{Perturbative reheating after
  multiple-field inflation: The impact on primordial observables},''
  \href{http://dx.doi.org/10.1103/PhysRevD.89.063535}{{\em \prd} {\bfseries 89}
  no.~6, (Mar., 2014) 063535}, \href{http://arxiv.org/abs/1311.3972}{{\ttfamily
  arXiv:1311.3972}}.

\bibitem{lindebook05}
A.~{Linde}, ``{Particle Physics and Inflationary Cosmology},'' {\em ArXiv High
  Energy Physics - Theory e-prints} (Mar., 2005) ,
  \href{http://arxiv.org/abs/hep-th/0503203}{{\ttfamily hep-th/0503203}}.

\bibitem{ichikawaetal08}
K.~{Ichikawa}, T.~{Suyama}, T.~{Takahashi}, {\em et~al.}, ``{Primordial
  curvature fluctuation and its non-Gaussianity in models with modulated
  reheating},'' \href{http://dx.doi.org/10.1103/PhysRevD.78.063545}{{\em \prd}
  {\bfseries 78} no.~6, (Sept., 2008) 063545},
  \href{http://arxiv.org/abs/0807.3988}{{\ttfamily arXiv:0807.3988}}.

\bibitem{kamadaetal11}
K.~{Kamada}, K.~{Kohri}, and S.~{Yokoyama}, ``{Affleck-Dine baryogenesis with
  modulated reheating},''
  \href{http://dx.doi.org/10.1088/1475-7516/2011/01/027}{{\em \jcap} {\bfseries
  1} (Jan., 2011) 027}, \href{http://arxiv.org/abs/1008.1450}{{\ttfamily
  arXiv:1008.1450 [astro-ph.CO]}}.

\bibitem{planck2015}
{\bfseries Planck} Collaboration, P.~A.~R. Ade {\em et~al.}, ``{Planck 2015
  results. XX. Constraints on inflation},''
  \href{http://dx.doi.org/10.1051/0004-6361/201525898}{{\em Astron. Astrophys.}
  {\bfseries 594} (2016) A20},
\href{http://arxiv.org/abs/1502.02114}{{\ttfamily arXiv:1502.02114
  [astro-ph.CO]}}.

\bibitem{Ade:2015tva}
{\bfseries BICEP2, Planck} Collaboration, P.~A.~R. Ade {\em et~al.}, ``{Joint
  Analysis of BICEP2/$Keck ?Array$ and $Planck$ Data},''
  \href{http://dx.doi.org/10.1103/PhysRevLett.114.101301}{{\em Phys. Rev.
  Lett.} {\bfseries 114} (2015) 101301},
\href{http://arxiv.org/abs/1502.00612}{{\ttfamily arXiv:1502.00612
  [astro-ph.CO]}}.

\bibitem{ForemanMackey:2012ig}
D.~Foreman-Mackey, D.~W. Hogg, D.~Lang, {\em et~al.}, ``{emcee: The MCMC
  Hammer},'' \href{http://dx.doi.org/10.1086/670067}{{\em Publ. Astron. Soc.
  Pac.} {\bfseries 125} (2013) 306--312},
\href{http://arxiv.org/abs/1202.3665}{{\ttfamily arXiv:1202.3665
  [astro-ph.IM]}}.

\bibitem{Ade:2015xua}
{\bfseries Planck} Collaboration, P.~A.~R. Ade {\em et~al.}, ``{Planck 2015
  results. XIII. Cosmological parameters},''
  \href{http://dx.doi.org/10.1051/0004-6361/201525830}{{\em Astron. Astrophys.}
  {\bfseries 594} (2016) A13},
\href{http://arxiv.org/abs/1502.01589}{{\ttfamily arXiv:1502.01589
  [astro-ph.CO]}}.

\bibitem{Lyth:1998xn}
D.~H. Lyth and A.~Riotto, ``{Particle physics models of inflation and the
  cosmological density perturbation},''
  \href{http://dx.doi.org/10.1016/S0370-1573(98)00128-8}{{\em Phys. Rept.}
  {\bfseries 314} (1999) 1--146},
\href{http://arxiv.org/abs/hep-ph/9807278}{{\ttfamily arXiv:hep-ph/9807278
  [hep-ph]}}.

\bibitem{Banks:2003sx}
T.~Banks, M.~Dine, P.~J. Fox, {\em et~al.}, ``{On the possibility of large
  axion decay constants},''
  \href{http://dx.doi.org/10.1088/1475-7516/2003/06/001}{{\em JCAP} {\bfseries
  0306} (2003) 001},
\href{http://arxiv.org/abs/hep-th/0303252}{{\ttfamily arXiv:hep-th/0303252
  [hep-th]}}.

\bibitem{Svrcek:2006yi}
P.~Svrcek and E.~Witten, ``{Axions In String Theory},''
  \href{http://dx.doi.org/10.1088/1126-6708/2006/06/051}{{\em JHEP} {\bfseries
  06} (2006) 051},
\href{http://arxiv.org/abs/hep-th/0605206}{{\ttfamily arXiv:hep-th/0605206
  [hep-th]}}.

\end{thebibliography}\endgroup


\newpage

\end{document}